\documentclass[11pt]{article}
\usepackage{authblk}
\usepackage{fullpage}
\usepackage[letterpaper, margin=1in]{geometry}
\usepackage{cite}
\usepackage{graphicx} 
\usepackage{setspace}
\usepackage{tikz}
\usetikzlibrary{arrows.meta,positioning}
\usepackage{enumitem}
\setlist{itemsep=1pt, topsep=3pt}
\usepackage{lmodern}
\usepackage{microtype}
\usepackage{needspace}

\title{Efficient Stable Word-RAM Simulation by Population Protocols}

\author[1]{Leszek Gąsieniec}
\author[2]{Tytus Grodzicki} 
\author[2]{Grzegorz Stachowiak}
\affil[1]{School of Computer Science and Informatics, University of Liverpool}
\affil[2]{Institute of Computer Science, University of Wroc{\l}aw }

\date{}

\usepackage{lineno}

\usepackage{amsfonts,amsmath,amssymb,amsthm}
\usepackage{tcolorbox}
\usepackage{multicol}
\usepackage{environ}
\usepackage[symbol]{footmisc}

\usepackage[T1]{fontenc}
\usepackage{graphicx}
\usepackage[dvipsnames]{xcolor}
\usepackage[utf8]{inputenc}
\usepackage{calc}
\usepackage{thmtools}
\usepackage{thm-restate}

\usepackage{booktabs}
\usepackage{tabularx}
\usepackage{array}

\newtheorem{theorem}{Theorem}
\newtheorem{lemma}{Lemma}

\newtheorem{definition}{Definition}

\newtheorem{example}{Example}
\newtheorem{fact}{Fact}

\newcommand{\mathescape}[1]{\relax\ifmmode#1\else $#1$\fi} 

\usepackage{algorithm,algorithmicx}
\usepackage[noend]{algpseudocode}
\makeatletter
  \renewcommand{\ALG@name}{Protocol}
\makeatother
\MakeRobust{\Call}
 
\newcommand{\LineIf}[2]{\State \algorithmicif\ {#1}\ \algorithmicthen\ {#2}}
\newcommand{\indLineIf}[2]{\State \hspace{\algorithmicindent} \algorithmicif\ {#1}\ \algorithmicthen\ {#2}}%
\newcommand{\DLineIf}[2]{\State \algorithmicif\ {#1}\ \algorithmicthen\ \State \hspace{\algorithmicindent} {#2}}
\newcommand{\LineIfElse}[3]{\State \algorithmicif\ {#1}\ \algorithmicthen\ {#2} \algorithmicelse\ {#3}}
\newcommand{\DLineIfElse}[3]{\State \algorithmicif\ {#1}\ \algorithmicthen\ {#2} \State \algorithmicelse\ {#3}}

\newcommand{\indindLineIfElse}[3]{\State \hspace{\algorithmicindent}\hspace{\algorithmicindent} \algorithmicif\ {#1}\ \algorithmicthen\ {#2} \State \hspace{\algorithmicindent}\hspace{\algorithmicindent} \algorithmicelse\ {#3}}
\renewcommand{\algorithmicrequire}{\textbf{Input:}}
\renewcommand{\algorithmicensure}{\textbf{Result:}}

\algdef{SE}[UNT]{Unt}{EndUnt}[1]{\textbf{until}\ #1\ \algorithmicdo}{\algorithmicend\ \textbf{until}}%

\definecolor{light-gray}{gray}{0.95} 
\usepackage{listings}
\lstdefinestyle{mystyle}{
    basicstyle=\ttfamily\small, 
    commentstyle=\color{gray},
    backgroundcolor=\color{light-gray},
    morecomment=[f][\color{grey}][0]{\#},
    morecomment=[f][\color{grey}][0]{//},
    numbers=left,                    
    numbersep=5pt,    
    mathescape=true,
    keywordstyle=\textbf,
    morekeywords={while,if,then,else,true,false,do,:,parallel,in,def,until},
}
\newcommand{\poly}{\operatorname{poly}}

\usepackage{xurl}
\usepackage[colorlinks=true,linkcolor=blue!60!black,citecolor=green!40!black,urlcolor=blue!60!black]{hyperref}

\DeclareTextFontCommand{\textsc}{\normalfont\scshape}

\begin{document}
\hypersetup{pageanchor=false}

\maketitle

\thispagestyle{empty}

\begin{abstract}
The \emph{population protocol} model studies computation by simple, indistinguishable \emph{agents} whose states evolve through pairwise \emph{interactions}. We consider the standard probabilistic variant, in which each interacting pair is selected uniformly at random. Inputs and outputs are represented in unary by the numbers of agents in designated states. Complexity is measured by the size of the state space and the \emph{stabilisation time}, defined as the number of interactions until stabilisation divided by the population size~$n$. A protocol is \emph{efficient} if both measures are polylogarithmic in~$n$, \emph{stable} if it eventually stabilises to the correct output with probability~$1$, and \emph{silent} if no agent changes state after stabilisation.

For fundamental problems such as leader election and majority, fast stable protocols in leaderless populations require a number of states that grows with~$n$. Despite substantial progress on these and other individual problems, including population size estimation, parity, and congruence, the general computational power of efficient stable population protocols remains largely unexplored.

We show that efficient stable population computation extends beyond individual problems and beyond the semilinear predicates, which characterise stable computation in the constant-state model. We develop a general framework for simulating $\mathcal{P}$-programs: word-RAM programs whose running time and number of registers are polynomial in the word length $w=\Theta(\log n)$. Our \textbf{main result} gives an efficient, stable, and silent simulation of every such program using only probabilistically scheduled pairwise interactions, with polylogarithmic stabilisation time both in expectation and with high probability (Theorem~\ref{t:main}). The framework supports general word-oriented computations, including multiple outputs, and is not restricted to decision problems. 
These results open a systematic study of efficient stable population protocols: their computational power, resource trade-offs, and extensions to external or spatial data.

Our construction extends the MC$+$ protocols and \emph{system of weights} introduced recently. MC$+$ protocols are fast Monte Carlo protocols whose failures, called \emph{anomalies}, are eventually detected. Using MC$+$ majority comparisons against the weights, the protocol converts unary inputs into binary registers, with each bit stored in an individual agent. It maintains and manipulates these registers, directly emulating each RAM instruction through MC$+$ operations on single bits, before converting the binary outputs back to unary. If an anomaly is detected, it switches to a new universal rank-based backup that simulates the program stably and silently.
\end{abstract}

\clearpage
\pagenumbering{arabic}
\hypersetup{pageanchor=true} 

\section{Introduction}

The population protocol model~\cite{AAD+06} studies computation by populations of simple,
indistinguishable entities, called \emph{agents}, that communicate through
\emph{pairwise interactions}. Each agent stores a single \emph{state} from a prescribed
\emph{state space}, and an interaction changes the states of the participating
agents according to a common deterministic \emph{transition function}. The \emph{initial configuration}
encodes the input, while the computation is required to \emph{stabilise} in a
configuration representing the solution.

We consider the standard \emph{probabilistic variant}, in which a \emph{random scheduler}
selects an ordered pair of distinct agents, called the \emph{initiator} and the
\emph{responder}, uniformly at random at each step. The \emph{sequential time} is the number of interactions until stabilisation, whereas the
\emph{parallel time}, or simply the \emph{time}, is the sequential time
divided by the population size~$n$. We call a protocol \emph{efficient} if it
uses a polylogarithmic number of states and stabilises in polylogarithmic
time. A protocol is \emph{stable} if it eventually stabilises to the
correct output with probability~$1$, and \emph{silent} if no agent changes
state after stabilisation.
An event occurs \emph{with high probability} (\emph{whp}) if its probability is at least $1-n^{-\eta}$ for any prescribed constant $\eta>0$, with the protocol constants chosen accordingly. Its complement has \emph{negligible} probability.
All logarithms are to base~$2$. Our simulation is stable and silent, with polylogarithmic stabilisation time both in expectation and whp.

We write a transition as
$A+B\longrightarrow A'+B',$
meaning that an interaction between an initiator in state $A$ and a responder
in state $B$ changes their states to $A'$ and $B'$, respectively. Some of our
protocols use \emph{compound states}: an agent state is a tuple of components
from sets $S_1,\ldots,S_k$, giving the full state space
$S_1\times\cdots\times S_k$.

\subsection{Related Work and Results}
\label{s:related}

Population protocols have been studied from two complementary perspectives:
which behaviours and predicates emerge from local rules, and how scheduling,
state space, agent capabilities, time, and computational power interact.
We primarily adopt the latter perspective, seeking general computations
that are simultaneously efficient, stable, and silent.
\smallskip

\noindent\textbf{Computational power of the classical model.}
In the original model~\cite{AAD+06}, the state space and transition function
are independent of~$n$. Under uniformly random interactions, these protocols
can simulate whp counter machines with a constant number of $O(n)$-bounded
unary counters represented by state multiplicities. Under unary encoding,
predicates computable with bounded error lie in $\mathsf{P}$ and admit randomised logarithmic-space
simulations with exponential running time~\cite{AAD+06}.
The stably computable predicates are exactly the semilinear predicates,
equivalently those definable in Presburger
arithmetic~\cite{AngluinAE06,DBLP:journals/dc/AngluinAER07}.
In the closely related model of chemical reaction networks, the stably
computable functions are exactly those with semilinear graphs~\cite{ChenDS14}.

These characterisations do not imply efficient stabilisation. With a
constant number of states, stable leader election~\cite{DBLP:conf/wdag/DotyS15}
and broad classes of predicates and functions under leaderless
initialisation~\cite{BDS17} require $\Omega(n)$ expected time.
For monotone protocol families, leader election and majority require
$\Omega(n/\operatorname{polylog} n)$ expected time with at most
$\frac12\log\log n$ states~\cite{DBLP:conf/soda/AlistarhAEGR17}.
Under the additional output-dominance assumption, majority in expected
time $O(n^{1-c})$, for any fixed $c>0$, requires $\Omega(\log n)$
states~\cite{DBLP:conf/soda/AlistarhAG18}.
\emph{Our growing state spaces support efficient stable computation beyond
the semilinear predicates that characterise constant-state stable computation.}
\smallskip

\noindent\textbf{Fast stable primitives.}
Basic coordination tasks become efficiently solvable when the number of states grows slowly with~$n$.
The first efficient leader-election protocol was given in~\cite{DBLP:conf/icalp/AlistarhG15}, and the number of states was reduced in~\cite{DBLP:conf/podc/BilkeCER17} and then to the optimal $O(\log\log n)$ in~\cite{GasieniecStachowiak18}, with $O(\log^2 n)$ time whp.
Further work~\cite{DBLP:conf/soda/BerenbrinkKKO18,DBLP:conf/spaa/GasieniecSU19,DBLP:journals/tpds/SudoOIKM20} culminated in~\cite{DBLP:conf/stoc/BerenbrinkGK20}, with $\Theta(\log\log n)$ states and optimal $O(\log n)$ expected time.
 
For exact majority, early fast
protocols~\cite{alistarh2015fastExactMajority,DBLP:conf/podc/BilkeCER17}
were followed by an $O(\log n)$-state, $O(\log^2 n)$-time protocol
based on a new leaderless phase clock~\cite{DBLP:conf/soda/AlistarhAG18}.
Further improvements~\cite{Berenbrink_2020,ben2020log3}
culminated in the stable protocol of~\cite{DBLP:conf/focs/DotyEGSUS21},
using $O(\log n)$ states and
optimal $O(\log n)$ expected time. With a sufficiently large initial
majority margin, approximate majority can be solved in $O(\log n)$ time
whp using only three states~\cite{DBLP:journals/dc/AngluinAE08}.
Efficient stable protocols for parity and congruence were obtained only
recently~\cite{gasieniec_et_al:LIPIcs.DISC.2026.17}.
 
Fast primitives have also been extended to \emph{graphical} population
protocols, where interactions follow an underlying graph: leader election
and exact majority on regular graphs, with overhead depending on
conductance~\cite{DBLP:conf/opodis/AlistarhGR21}; near-optimal stable leader
election on general graphs~\cite{DBLP:conf/podc/AlistarhRV22}; and
space-efficient exact majority on general graphs~\cite{Ryb2}.
\emph{Our simulation assumes a complete graph; extending it to these
interaction graphs remains open.}
 
Many of these protocols use the \emph{one-way epidemic
process}~\cite{DBLP:journals/dc/AngluinAE08a}, whose agents are either susceptible~($0$)
or infected~($1$), with transition
$(x,y)\rightarrow(x,\max\{x,y\})$.
\begin{lemma}[Lemma~2 in~\cite{DBLP:journals/dc/AngluinAE08a}]
Starting from a single infected agent, the epidemic reaches all agents in
$\Theta(\log n)$ time whp.
\end{lemma}
This dissemination primitive underlies phase clocks and many fast
leader-election and majority protocols; see~\cite{GKL+24} for a
$k$-contact generalisation.
 
Another key ingredient of these protocols is synchronisation.
Phase clocks coordinate successive stages of computation, typically
allowing $\Theta(\log n)$ time for epidemic dissemination.
Given a leader or a suitably sized junta, constant-state
clocks~\cite{DBLP:journals/dc/AngluinAE08a,DBLP:journals/jacm/GasieniecS21}
can be nested to measure any fixed polylogarithmic duration.
Leaderless constructions use $\Theta(\log n)$ states, including
fixed-resolution clocks~\cite{DBLP:conf/focs/DotyEGSUS21} and clocks
based on the power-of-two-choices process~\cite{DBLP:conf/soda/AlistarhAG18}.
A different leader-based construction measures $\Theta(n\log n)$ time
whp using a constant number of states~\cite{DBLP:conf/stacs/GasieniecSS23}.

\emph{We combine MC$+$ leader election and majority~\cite{gasieniec_et_al:LIPIcs.DISC.2026.17},
whose failures are eventually detected, with epidemics and a leader-based clock to
simulate general programs; faster MC$+$ primitives could sharpen our bounds.}
 
Fast protocols also support counting and computing proportions~\cite{MocquardAABS2015,mocquard2016optimal}.
Uniform population-size estimation has also been
studied~\cite{DotyEftekhari19}, including stable protocols that compute
either $\lfloor\log n\rfloor$ or $\lceil\log n\rceil$ in
$O(\log^2 n)$ time whp~\cite{DBLP:conf/podc/BerenbrinkKR19};
see also~\cite{DOTY202191} for a survey of counting.
\emph{Our simulation assumes known $w=\Theta(\log n)$; stable size
estimation may help remove this assumption.}
\smallskip

\noindent\textbf{Fast register-machine and program simulations.}
A close predecessor is the register-machine simulation
in~\cite{DBLP:journals/dc/AngluinAE08a}. Given an initial leader, it simulates
a machine with a constant number of registers whp,
supporting comparison, addition, subtraction, and multiplication or division
by constants, with $O(\log^2 n)$ time per operation after initialisation.
The same machinery simulates a logarithmic-space Turing machine.
The registers are represented by state multiplicities, rather than
explicit binary words, and the general simulation permits a small
probability of error. A separate construction in the same work guarantees
correct convergence with probability~$1$ for semilinear predicates.
 
In stochastic chemical reaction networks, a fixed finite reaction network
can perform universal computation with arbitrarily small error probability
and polynomial slowdown~\cite{SoloveichikCWB08}.
 
The framework of~\cite{DBLP:conf/podc/KosowskiU18} uses nested phase clocks
and composition to support a restricted imperative language with loops and branching. Every semilinear predicate
admits a constant-state protocol with expected polylogarithmic convergence
time and high-probability correctness. Its always-correct variants
achieve expected $O(n^{\varepsilon})$ convergence time for any fixed
$\varepsilon>0$, with a constant number of states depending on
$\varepsilon$, or retain polylogarithmic convergence time using
$O(\log\log n)$ states. These guarantees concern convergence; the
always-correct constructions are not silent as presented. Convergence concerns
the realised execution; stabilisation rules out every reachable output change.
\emph{With growing state spaces, we simulate general word-RAM programs on
binary registers, combining probability-one correctness and silence with
polylogarithmic time in expectation and whp.}
\smallskip

\noindent\textbf{Presburger computation and succinct protocols.}
For a quantifier-free Presburger formula of size~$m$ as defined in~\cite{DBLP:journals/jcss/CzernerGHE24}, succinct protocols~\cite{blondin_et_al:LIPIcs.STACS.2020.40} reduce exponential state complexity to $\poly(m)$, at the cost of exponential expected time, as observed in~\cite{DBLP:journals/jcss/CzernerGHE24}.
Using population computers as an intermediate formalism, the construction
of~\cite{DBLP:journals/jcss/CzernerGHE24} achieves $O(m)$ states and
$O(m^7 n)$ expected time for $n=\Omega(m)$. Linear time is worst-case optimal for leaderless constant-state protocols,
whose predicates are semilinear.
\emph{For leaderless stable protocols, we propose a relaxed notion of succinctness allowing $\poly(m)\cdot\poly\log n$ states, which our framework achieves with silent stabilisation (Section~\ref{s:conclusions}).}
\smallskip

\noindent\textbf{Expressive power with growing state spaces.}
Uniform protocols with $o(\log n)$ states still decide only semilinear
predicates~\cite{DBLP:journals/dc/AngluinAER07,DBLP:journals/tcs/ChatzigiannakisMNPS11},
whereas protocols with $f(n)\in\Omega(n)$ states capture
$\mathsf{NSPACE}(n\log f(n))$ under unary
encoding~\cite{DBLP:journals/tcs/ChatzigiannakisMNPS11}.
For the intermediate range, under the constructibility and uniformity
assumptions of~\cite{CzernerFischerGuttenberg25},
for $f(n)\in\Omega(\log n)\cap O(n^{1-\varepsilon})$ and fixed
$\varepsilon>0$, uniform and weakly uniform protocols with
$\Theta(f(n))$ states decide exactly the predicates whose unary encodings
lie in $\mathsf{NSPACE}(f(n)\log n)$.
The construction simulates space-bounded nondeterministic machines through
leader-driven counter machines and restarts, potentially providing a backup.
\emph{The characterisation ensures correctness under fairness, leaving
random-scheduler time and silence unanalysed; our word-RAM backup supplies
polynomial expected time and silence.}
\smallskip

\noindent\textbf{Simulating population protocols.}
Under the uniformity assumptions above, polylogarithmically many states
permit only predicates in $\mathsf{NSPACE}(\operatorname{polylog} n)$,
even without a time bound. A complementary time bound follows by simulating
protocols on conventional machines, provided their descriptions can be
generated in $\operatorname{poly}(w)$ time.
A randomised $w$-RAM stores one $O(w)$-bit count per state and samples
interactions. For a silent protocol with expected polylogarithmic
stabilisation time, it detects termination by checking every feasible pair
of states for a transition that changes either agent's state.
Since $w=\Theta(\log n)$, the expected number of interactions is $n\operatorname{polylog}n$. They can be simulated by a Las Vegas algorithm using $\poly(w)$ space and expected $2^{O(w)}$ time.
Batched simulations~\cite{BerenbrinkHK0PT20,HintzeP26}
reduce the amortised cost per interaction, but do not by themselves yield a
general $\poly(w)$-time simulation.
\emph{With binary counts as RAM inputs, our $\poly(w)$-time source
computations leave a gap to this expected exponential-time reverse
simulation; narrowing it remains open.}
\smallskip

\noindent\textbf{More powerful extensions of population protocols.}
Several extensions strengthen agent resources or communication.
Community protocols~\cite{DBLP:conf/icalp/GuerraouiR09} give agents unique identifiers
and allow them to remember a constant number of identifiers, capturing
symmetric predicates in $\mathsf{NSPACE}(n\log n)$.
Homonym protocols~\cite{DBLP:journals/mst/BournezCR18} allow shared identifiers; with the
model's identifier operations, polylogarithmically many distinct identifiers
suffice to simulate Turing machines using polylogarithmic space.
PALOMA~\cite{DBLP:journals/tcs/ChatzigiannakisMNPS11} gives agents logarithmic local memory
and captures the same class as community protocols.
 
Broadcast consensus protocols capture unary predicates in
$\mathsf{NL}$~\cite{BlondinEsparzaJaax19}.
The simulation of~\cite{CzernerJaax21} emulates a randomised
logarithmic-space Turing machine of expected running time~$T$ in expected
$O(n\log n\cdot T)$ interactions.
Clocked population protocols~\cite{Aspnes2021} provide an idealised signal
that sufficient time for convergence has elapsed; absence
detectors~\cite{DBLP:journals/jpdc/MichailS15} test whether a specified state is absent; and
selective population protocols~\cite{GanczorzGJKS24} strengthen partner
selection. \emph{Our simulation retains anonymous agents and uniformly random
pairwise interactions, without these additional capabilities.}
\smallskip
 
\noindent\textbf{Complexity-theoretic perspective.}
Stochastic well-structured transition systems~\cite{Asp25} encompass
population protocols, chemical reaction networks, and gossip models. Their clocks stop or advance too quickly after polynomially many expected steps, and explicitly terminating computations finish or fail in expected polynomial time.
Under bounded-error termination semantics, the unaugmented models capture
the symmetric languages in $\mathsf{BPL}$.
\emph{These fixed-state results concern bounded-error termination; our
construction uses polylogarithmically many states and clock rounds, with a
clock-independent backup ensuring probability-one correctness.}
\smallskip

\noindent\textbf{Stability, silence, and the MC$+$ paradigm.}
Stable computation guarantees an eventually correct, unchanging output;
silence additionally forbids subsequent state changes. The class of well-specified strongly silent
protocols introduced in~\cite{BlondinEJM21} preserves the
classical expressive power and supports more tractable automatic
verification. Silence is natural for machine simulation: once the program
terminates and its output is decoded, no clock or synchronisation mechanism
should continue modifying the population.
 
The framework in~\cite{gasieniec_et_al:LIPIcs.DISC.2026.17} combines MC$+$
protocols and population weights. A fast probabilistic computation detects
violations of its correctness conditions within polynomial expected time. These violations, called
\emph{anomalies}, are absent whp; when one is detected, the protocol
switches to an always-correct backup. This yielded efficient, stable, and
silent protocols for parity and congruence.

\emph{We extend this framework to general word-RAM simulation, reducing
the leader's clock component after initialisation by $\Theta(\log n)$,
creating weights in one round, and providing unary--binary conversions
for multiple inputs and outputs.}
\smallskip

\noindent\textbf{Our contribution.}
The preceding results leave a gap between fast but Monte Carlo
register-machine simulations with $O(1)$ registers, efficient stable protocols for individual
problems, characterisations of growing-state protocols without time bounds, and
machine simulations available only in stronger variants of the model.
We close part of this gap and thereby address the time-bounded expressive power of growing-state protocols,
identified as an open problem in~\cite{CzernerFischerGuttenberg25}.
 
We consider $\mathcal{P}$-programs (Definition~\ref{P-programs}):
$w$-RAM programs whose running time and number of registers are polynomial in $w=\Theta(\log n)$, which halt on every admissible input and initialise each working register before its first read. Our \textbf{main result} gives an efficient, stable, and silent
simulation of every such program, with polylogarithmic time both in
expectation and whp (Theorem~\ref{t:main}), and thus a lower bound on the computational power of efficient stable protocols.
The simulation supports general word-oriented
computations, including multiple numerical outputs, and thus contributes to
the study of computation beyond agreement identified
in~\cite{DBLP:books/sp/09/AspnesR09}.
 
Applications include functions with non-semilinear graphs, such as products
of input counts when the result fits the prescribed output range.
Section~\ref{s:conclusions} discusses further applications and extensions: succinct protocols for quantifier-free Presburger formulas,
following~\cite{DBLP:journals/jcss/CzernerGHE24}, and a possible extension to expected-polynomial-time Las Vegas computations.

We retain the unary input and output representation standard for population protocols, while RAM instructions operate on binary words.
Our main technical contribution is an MC$+$ implementation of the entire
fast computation (Section~\ref{s:MC$+$_sim}), including unary--binary conversion
and the maintenance and modification of binary registers.
Majority comparisons against the population weights convert unary inputs
into binary registers, each represented by $w$ agents, one per bit.
The protocol emulates each RAM instruction through $O(w)$ atomic subroutines that read or write individual bits, then converts the binary outputs back to unary form. If an anomaly is detected, a novel rank-based backup simulates the program stably and silently in expected $O(n^2w)$ time (Section~\ref{s:stable_sim}).
Together, the MC$+$ computation and backup guarantee correct silent
stabilisation with probability~$1$, using polylogarithmically many states.

\section{Model and Simulation Overview}\label{s:model}\label{app:section2}

Population protocols represent numerical inputs and outputs in unary: input $A_i$ is represented by a set $X_i$ of $|X_i|=A_i$ agents, and output $B_i$ by a set $Y_i$ of $|Y_i|=B_i$ agents. Outputs need not be Boolean; leader election, for example, produces a singleton output set. Values can be as large as the population size~$n$, so their binary representations have logarithmic length. The word-RAM~\cite{CookR73, DBLP:conf/stoc/FredmanW90}, with $w=\Theta(\log n)$, therefore matches their scale and provides a standard model of general-purpose computation in which to establish efficient, stable, and silent population simulation.
\smallskip

\noindent\textbf{The $w$-RAM model.}
The basic parameter of the $w$-RAM model is its word length~$w$.
More specifically, (1)~its instruction set is based on the Cook--Reckhow model~\cite{CookR73} (Table~\ref{tab:c-rram}), (2)~each input and output value is a non-negative $(w-1)$-bit integer, and (3)~registers store signed $w$-bit words in two's-complement (U2) encoding. Arithmetic is performed modulo~$2^w$.
The first six instructions listed in Table~\ref{tab:c-rram} are referred to as
\emph{RAM instructions}, whereas $\Call{READ}{}$ and $\Call{PRINT}{}$ are
referred to as \emph{input/output instructions}.

\begin{table}[H]
\centering
\begin{tabular}{l||l}
\textbf{Instruction} & \textbf{Description} \\\hline\hline
$\Call{LOAD}{C, rd}$ & $C \rightarrow rd$: load an integer constant $C$ into register $rd$ \\
$\Call{ADD}{rs1, rs2, rd}$ & $[rs1] + [rs2] \rightarrow rd$: load the sum of the values in $rs1, rs2$ into $rd$\\
$\Call{SUB}{rs1, rs2, rd}$ & $[rs1] - [rs2] \rightarrow rd$: load the difference of the values in $rs1, rs2$ into $rd$\\
$\Call{COPY}{i, rp, d, rd}$ & $[[rp]] \rightarrow rd$: load the value of the register pointed to by $rp$ into $rd$\\
$\Call{COPY}{d, rs, i, rp}$ & $[rs] \rightarrow [[rp]]$: store the value of $rs$ in the register pointed to by $rp$\\
$\Call{JGE}{r, Iz}$ & if $[r] \ge 0$, jump to instruction $Iz$; otherwise continue \\\hline
$\Call{READ}{rd}$ & read the next input value into register $rd$\\
$\Call{PRINT}{rs}$ & output the value of register $rs$\\
\end{tabular}
\caption{Instructions of the Cook--Reckhow RAM model.\protect\footnotemark}
\label{tab:c-rram}
\end{table}
\footnotetext{The square brackets $[\cdot]$ denote the value stored in a register.}

A \emph{$w$-RAM program} is a finite sequence of instructions from
Table~\ref{tab:c-rram}. Its execution is controlled by an
\emph{instruction pointer} $ip$, which by default advances to the next
instruction and may be modified explicitly by a conditional jump $\Call{JGE}{}$. Input and
output are handled by the corresponding input/output instructions.
The execution terminates when $ip$ reaches the end of the program. 
Although the number of registers in the standard word-RAM model
is not usually fixed in advance, in this paper we restrict our attention to
$w$-RAM programs whose memory consists of $\poly(w)$ registers, identified
with their indices $0,1,\ldots,\rho-1$.
\footnote{A program using sparse
addresses can be transformed to this form by storing the addresses in use in a
binary trie, which increases its running time by a factor polynomial in~$w$.}
More precisely, the class of programs considered in this paper is captured by
the following definition of $\mathcal{P}$-programs.

\begin{definition}[$\mathcal{P}$-programs]\label{P-programs}
A $w$-RAM program is called a $\mathcal{P}$-program if it satisfies the
following properties:
\begin{enumerate}
    \item the total number of instructions executed during its runtime is at
    most
    \[
        T=T_R+T_{\mathrm{RAM}}+T_P=\poly(w),
    \]
    where $T_R$ and $T_P$ denote the numbers of executed $\Call{READ}{}$ and
    $\Call{PRINT}{}$ instructions, and $T_{\mathrm{RAM}}$ is an upper bound
    on the number of executed RAM instructions;
    \item the program uses $\rho=O(T)=\poly(w)$ registers, indexed from $0$ to
    $\rho-1$;
    \item on every admissible tuple of $T_R$ input values, the program halts within these bounds, reads exactly $T_R$ values, and produces exactly $T_P$ non-negative outputs smaller than $2^{w-1}$;
    \item on every execution, each working register is written before its first read, including a read used for indirect addressing, and every direct or indirect register access lies in $\{0,\ldots,\rho-1\}$.
\end{enumerate}
\end{definition}

These requirements also apply to the partial input counts used during backup retries. The program therefore restarts independently of old working-register contents; its registers are disjoint from the arrays and auxiliary registers added by the simulation. We take $T_{\mathrm{RAM}}\ge1$, increasing this upper bound if necessary.

This class encompasses a broad range of computations.
Thus, the ability to simulate it shows that silently stabilising population
protocols can efficiently solve a wide range of computational problems.
Establishing this result is the main contribution of this paper, as stated in
Theorem~\ref{t:main}.

\smallskip

\noindent\textbf{Population size and data representation.}
Each input and output variable is represented in unary.
For a given $\mathcal{P}$-program, the simulation uses a distinct input state
for each input variable and a distinct output state for each output variable,
namely
$
    S_{\mathrm{in}}=\{in_1,\ldots,in_{T_R}\},
    S_{\mathrm{out}}=\{out_1,\ldots,out_{T_P}\},
$
so that $|S_{\mathrm{in}}|+|S_{\mathrm{out}}|=T_R+T_P=\poly(w)$.
Each input and output set has fewer than $2^{w-1}$ agents.
In contrast, the $w$-RAM model uses binary representations of numbers, which
we also found convenient for simulating $w$-RAM computations.
Our simulation therefore converts the input values from unary to binary before
emulating the $\mathcal{P}$-program, and converts the final output values back to unary
after the computation has been completed.
Thus, the external input and output remain in unary form, while all
intermediate computations use binary representations.
The input values are stored in an \emph{input array} of $T_R$ registers, and
the results are collected in an \emph{output array} of $T_P$ registers before
being converted to unary. Together with $O(1)$ auxiliary registers, the total
number of registers in the simulation is therefore
$
    \rho'=\rho+T_R+T_P+O(1)=\poly(w).
$
The total binary input length is $O(T_Rw)$ and need not equal~$w$.

During the simulation, the population must accommodate the unary input and
output values, the \emph{system of weights}, which consists of $2^{w-1}-1$ agents (see Subsection~\ref{ss:MC$+$_clock_le}), and the
$\rho'w$ agents -- each representing one bit of a register. In addition, a constant fraction of
all agents should remain free and form the \emph{working set}, which
allows the epidemic-based computations to run fast.
For these reasons, the simulation uses a population of size
$
    n_w=C\bigl((T_R+T_P+1)2^w+\rho'w\bigr),
$
where $C>1$ is a fixed integer constant.
The non-free roles are input, output, weight and register agents, the leader and, in the backup, a helper. A single simulation uses fewer than $n_w/C$ such agents.

\begin{fact}\label{f:free_agents}
For $n\ge n_w$, at least $(1-1/C)n$ agents are free whenever only one simulation's auxiliary roles are present, subject to the prescribed register and output bounds.
\end{fact}

This covers successful MC$+$ executions and the final backup after lower-ranked auxiliary states have been reclaimed. 
Since $T_R$, $T_P$, and $\rho'$ are $\poly(w)$, the following relations hold.

\begin{fact}\label{f:word_population_relation}
We have $\log n_w=w+O(\log w)=\Theta(w)$ and, in particular, $\log n_w<2w$ for all sufficiently large~$w$.
\end{fact}


For sufficiently large $w$, all $\rho'=\poly(w)$ register addresses fit in non-negative $w$-bit words. Our main result is the following theorem.

\begin{theorem}\label{t:main}
Every $\mathcal{P}$-program admits an efficient simulation by a silently stabilising probabilistic population protocol.
For all sufficiently large $w$ and population size
$n_w$, the simulation stabilises, both in expectation and with high
probability, in
$
O\!\left(
    w^2\,
    (T_R w + T_{\mathrm{RAM}} + T_P)
\right)
$
time, using
$
O\!\left(
    \rho' w^2\,
    (T_R w + T_{\mathrm{RAM}} + T_P)
\right)
$
states.
\end{theorem}

\begin{proof}
By Theorem~\ref{t:MC$+$sim}, there is an efficient MC$+$ simulation protocol that completes whp in $O(w\,(T_Rw+T_{\mathrm{RAM}}+T_P))$ rounds, each lasting $\Theta(\log n)=\Theta(w)$ time, plus one register initialisation round of $O(\rho'w)$ time, and uses $O\!\left(\rho'w^2\,(T_Rw+T_{\mathrm{RAM}}+T_P)\right)$ states.
By Theorem~\ref{t:stable_sim}, there is a silently stabilising backup simulation protocol with expected time $O(n^2w)$ using $O(\rho'w^2)$ states.
The fast protocol has a fixed round budget and retains anomaly checks through terminal verification. Subsection~\ref{ss:MC$+$_clock_le} gives a polynomial bound $D(n)$ on the expected remaining time, from any reachable configuration, to completion or detected failure followed by switching and backup. Thus, for the high-probability bound $A=O(\rho'w+w^2(T_Rw+T_{\mathrm{RAM}}+T_P))$, choosing the failure exponent $\eta$ sufficiently large gives expected time at most $A+n^{-\eta}D(n)=O(A)$.
By Theorem~\ref{th:stable+MC$+$}, the one-way combination is therefore stable and silent, with the same time bound whp. Its disjoint fast and backup state sets give the stated state bound. Finally, $\rho'=O(T)$ and $T\le T_Rw+T_{\mathrm{RAM}}+T_P$, so register initialisation is absorbed in the stated time bound.
\end{proof}

The protocol requires the $\mathcal{P}$-program, its word length~$w$, and the parameters $T_R$, $T_{\mathrm{RAM}}$, $T_P$, and $\rho$.
The analysis uses only $w=\Theta(\log n)$, which holds for $n=n_w$ by Fact~2.
It therefore applies without change to every population size $n\ge n_w$ with $\log n=\Theta(w)$, since additional agents are free and only increase the working set; the constants in the subprotocol durations depend on the constant hidden in $\log n=\Theta(w)$.
For significantly larger populations, the simulation remains efficient and stable provided that it is also given a constant-factor upper bound on $\log n$ for timing its subprotocols.
In the remainder of this section, we present the components and techniques
common to both variants of the simulation.

\subsection{General Simulation Scheme}\label{ss:general_scheme}

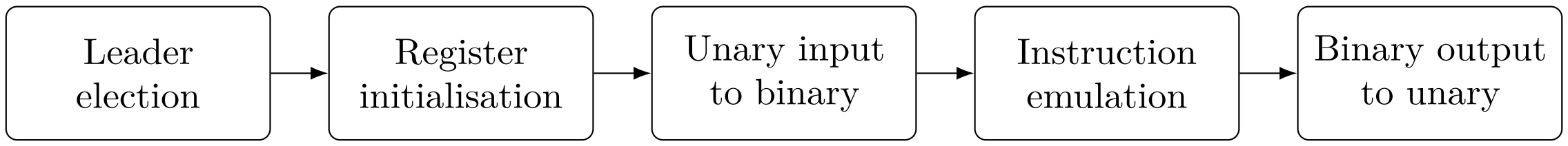
\begin{figure}[H]
    \centering
    \begin{tikzpicture}[
  stage/.style={draw,rounded corners=3pt,align=center,text width=2.25cm,
                minimum height=1.28cm,inner sep=4pt,font=\small},
  >=Latex,node distance=0.55cm]
\node[stage] (leader) {Leader\\election};
\node[stage,right=of leader] (registers) {Register\\initialisation};
\node[stage,right=of registers] (input) {Unary input\\to binary};
\node[stage,right=of input] (ram) {Instruction\\emulation};
\node[stage,right=of ram] (output) {Binary output\\to unary};
\draw[->] (leader) -- (registers);
\draw[->] (registers) -- (input);
\draw[->] (input) -- (ram);
\draw[->] (ram) -- (output);
\end{tikzpicture}
    \caption{High-level structure of the $\mathcal{P}$-program simulation.}
    \label{fig:sim_scheme}
\end{figure}

Each $\mathcal{P}$-program simulation, whether the efficient MC$+$
simulation or the stable backup, follows the same high-level scheme,
illustrated in Figure~\ref{fig:sim_scheme}.

\subsubsection{The Leader}\label{app:the_leader}
Both simulations begin with leader election. The leader coordinates every subsequent stage: initialising the registers, converting unary inputs to binary in the input array, acting as the processor during instruction emulation, and converting the results in the output array back to unary.

The simulated $\mathcal{P}$-program has fixed length, with execution determined by its input. Its code, including constants and direct register addresses, is encoded in the transition function. The leader maintains an instruction pointer $ip$ and coordinates the emulation of the current instruction. The fixed code requires only $O(1)$ states for $ip$ and its direct addresses (Example~\ref{ex:transition_rule}, page~\pageref{ex:transition_rule}). After the last instruction, $ip$ is set to $\mathsf{END}$.

The following example illustrates how an instruction of the program, determined by the leader's $ip$, is realised by transition rules that additionally depend on auxiliary state components.
\begin{example}\label{ex:transition_rule}
Suppose that register~$r$ is represented by a single agent storing its value, that instruction~$7$ sets this value to~$\square$, and that the update requires the leader and the register agent to share an auxiliary component, say the colour \emph{red}.
Since $r$ and $\square$ are determined by $ip$, the leader does not store them; the transition function contains the rule
\[
    (L;\mathit{red};\,ip{=}7) + (r;\mathit{red};\,v) \rightarrow
    (L;\mathit{red};\,ip{=}8) + (r;\mathit{red};\,\square),
\]
for every value~$v$.
Interactions in which the colours differ leave both agents unchanged.
\end{example}

\smallskip

\subsubsection{Registers}\label{app:reg_init}
Each of the $\rho'$ registers is represented by $w$ distinct agents in states $(R;r,i,b)$, where $r\in\{0,\ldots,\rho'-1\}$ is the register address, $i$ the bit index, and $b$ its value. Both simulations use the same initialisation (Protocol~\ref{alg:reg_init}, page~\pageref{alg:reg_init}): the leader assigns successive free agents to successive register bits, setting every bit to~$0$.

\begin{algorithm}[H]
\begin{algorithmic}[1]
\Require{A unique leader, the free-agent bound of Fact~\ref{f:free_agents}, $\rho'$, and $w$.}
\Ensure{Initialised registers indexed by $0,1,\ldots,\rho'-1$, each consisting
of $w$ bits.}
\Function{RegInit}{ }
\For{$r \gets 0$ \textbf{to} $\rho'-1$}
    \For{$i \gets 0$ \textbf{to} $w-1$}
        \State The leader assigns the $i$-th bit of register $r$ to the first
        free agent it encounters:
        \State \hspace{\algorithmicindent}
        $L+\varphi\rightarrow L+(R;r,i,0)$.\Comment{$\varphi$ denotes a free agent}
    \EndFor
\EndFor
\EndFunction
\end{algorithmic}
\caption{$\Call{RegInit}{ }$}\label{alg:reg_init}
\end{algorithm}

The time and correctness guarantees are proved with Lemma~\ref{l:reg_init} below.

\begin{lemma}\label{l:reg_init}
Assuming a unique leader and the free-agent bound of Fact~\ref{f:free_agents}, Protocol~\ref{alg:reg_init} stably and silently
initialises $\rho'$ registers of $w$ bits each, in $O(\rho'w)$ time both in
expectation and whp.
\end{lemma}

\begin{proof}\label{app:l:reg_init}
Put $M=\rho'w$ and $p=1-1/C$. Until allocation finishes, each interaction assigns the next bit with probability at least $2p/n$. Its duration is therefore stochastically dominated by the time needed for $M$ successes in Bernoulli trials with success probability $2p/n$, giving expected time at most $M/(2p)=O(M)$. After $dnM$ interactions, a Chernoff bound gives fewer than $M$ successes with probability $e^{-\Omega(M)}=n^{-\Omega(1)}$, for a sufficiently large constant $d$, since $M\ge w=\Theta(\log n)$. Every assignment removes its agent from the free set, so all register-bit agents are distinct. No allocation rule applies after the last assignment, giving stable and silent completion.
\end{proof}
\smallskip

\subsubsection{Input and Output Conversion}\label{app:io_conversion}
The $i$-th input variable, with value $k_i<2^{w-1}$, is given in unary by $k_i$ agents in state $in_i$; before emulating RAM instructions, the protocol converts it to binary and stores it in register $r^{in}_i$ of an \emph{input array} occupying a fixed range of consecutive addresses, with an additional pointer register $rp_{in}$.
Every $\Call{READ}{rd}$ is replaced by $\Call{COPY}{i,rp_{in},d,rd},$ $\Call{INC}{rp_{in}}$ ($\Call{INC}{}$ increments a register, see Subsection~\ref{ss:ram_instructions}), and analogously every $\Call{PRINT}{rs}$ by $\Call{COPY}{d,rs,i,rp_{out}},$ $\Call{INC}{rp_{out}}$, which writes to an \emph{output array} $r^{out}_1,\ldots,r^{out}_{T_P}$ with pointer $rp_{out}$; the pointers are initialised by two $\Call{LOAD}{}$ instructions as the arrays have fixed addresses.
The resulting \emph{modified} $\mathcal{P}$-program still has fixed length, performs no input/output instructions and executes at most $T_{\mathrm{RAM}}+2(T_R+T_P)+2$ RAM instructions; only after it has been emulated are the values in the output array converted from binary to unary.
Jump targets are adjusted to the expanded code. Within the modified program, only these trusted replacements access the reserved arrays. At each backup retry, the pointer loads are repeated, while the cumulative input array is preserved and the source program initialises its working registers. This preprocessing is essential for the backup.
Conversions use a system of weights in the MC$+$ simulation and controlled rewriting rules in the backup; the full description follows below.

The following paragraphs give the full description of the input and output conversion scheme used in both simulation variants.
\bigskip

Before the actual simulation of the $\mathcal{P}$-program, the protocol
converts the input data from unary to binary.
To store and access the converted input values, we use $T_R+1$ auxiliary
registers: an \emph{input array} $r^{in}_1,r^{in}_2,\ldots,r^{in}_{T_R}$ and an
additional register $rp_{in}$ containing a pointer to the next entry of the
input array.
The input array occupies a fixed range of consecutive addresses, with
$r^{in}_i$ at its $i$-th position.
The $i$-th integer input variable, with value $k_i<2^{w-1}$, is represented in
unary by a subpopulation of $k_i$ agents in state $in_i$.
After the conversion, its binary representation is stored in register
$r^{in}_i$.
Once all input data has been converted and stored in the input array, we
replace every instruction $\Call{READ}{rd}$ of the $\mathcal{P}$-program with
the pair of instructions
$
    \Call{COPY}{i,rp_{in},d,rd},\; 
    \Call{INC}{rp_{in}}.
$
The first copies the input-array entry pointed to by $rp_{in}$ into register
$rd$, and the second advances the pointer.
To make $rp_{in}$ initially point to the first entry of the input array,
the modified program starts with $\Call{LOAD}{r_1^{in},rp_{in}}$.
The $\Call{INC}{r}$ instruction increments the value of register~$r$ by one; it is
not an instruction of Table~\ref{tab:c-rram}, but it is emulated in the same
way as $\Call{ADD}{}$, see Subsection~\ref{ss:ram_instructions}.
This preprocessing is essential for our backup simulation.
Similarly, instead of performing binary-to-unary conversion on the fly during
printing, we place the printed values in consecutive entries of the
\emph{output array} $r^{out}_1,r^{out}_2,\ldots,r^{out}_{T_P}$, which also
occupies a fixed range of consecutive addresses, by replacing
every instruction $\Call{PRINT}{rs}$ with
$
    \Call{COPY}{d,rs,i,rp_{out}},\; 
    \Call{INC}{rp_{out}}.
$
Analogously, this requires $T_P+1$ auxiliary registers, including the pointer
register $rp_{out}$, which is initialised by $\Call{LOAD}{r_1^{out},rp_{out}}$ at the start of the program.

We thus obtain a RAM program, referred to as the \emph{modified} ${\cal P}$-program,
that retrieves its data from the input array, processes it in binary, and
stores the results in the output array. Its code is longer than that of the
original program by one instruction per $\Call{READ}{}$ and
$\Call{PRINT}{}$ instruction in the code, plus the two initial $\Call{LOAD}{}$ instructions, so it still has fixed length, and
during its execution it performs no input/output instructions and at most
$T_{\mathrm{RAM}}+2(T_R+T_P)+2$ RAM instructions. Jump targets are updated to the corresponding expanded source instructions, beyond the initial pointer loads. Within the modified program, only the trusted replacements access the reserved arrays. On a backup retry, the cumulative input array is retained, the two pointer loads are repeated, and the source program initialises its working registers. Since every tentative input tuple is admissible, the program halts and overwrites all $T_P$ output entries.
The computations of the modified program are emulated by a population protocol
that operates between the creation of the input array and the conversion of the
output array.
Only after the entire modified program has been emulated are the values in the
output array converted from binary to unary.
This scheme is used in both the efficient MC$+$ simulation and the stable
backup simulation.

We use two mechanisms for conversion between unary and binary representations.
The first is based on a system of weights and is employed in the efficient
MC$+$ simulation.
The second relies on controlled rewriting rules and is used in the stable and
silent backup simulation.

\smallskip

\subsubsection{State Organisation}
Table~\ref{tab:state_organisation}, page~\pageref{tab:state_organisation}, summarises the roles and state components described throughout the paper. We write $(a|b)$ for either state $a$ or state $b$. Only input agents have a permanent role, retaining their labels $in_i$. All other agents are initially leaders; after election, all but one are free~($\varphi$), forming the working set. Free agents subsequently become register agents~($R$), output agents, weight agents~($P$, MC$+$ only), or helpers~($H$, backup only). Every agent also holds a clock component in MC$+$ or a rank in the backup.

\begin{table}[H]
    \centering
    \begin{tabular}{l||l||l}
          \textbf{Role} & \textbf{Components} & \textbf{Used in} \\
         \hline\hline
           input (\textbf{permanent} role)& input state $in_i$ & both \\
         \hline\hline
          leader ($L$) & $ip$, register and bit index, $O(1)$ memory & both \\
          register $(R)$& $(r,i,b)$: register index, bit index, bit value & both \\
          output & output state $out_i$ & both \\
          free ($\varphi$) & request/response components of GET/PUT & MC$+$ \\
          weight ($P$) & weight index $k$ of $P_k$ & MC$+$ \\
          helper ($H$) & counter of interactions with the leader & backup \\
         \hline\hline
          every agent & clock component: round $t$, phase $x$ & MC$+$ \\
          every agent & rank & backup \\
    \end{tabular}
    \caption{Roles and main state components of agents. The input role is the only permanent one; the other roles are assigned to non-input agents during the computation.}
    \label{tab:state_organisation}
\end{table}

\subsection{RAM Instruction Emulation}\label{ss:ram_instructions}\label{app:reductions}

A key feature of our construction is that the emulation of every RAM instruction reduces to two \emph{atomic subroutines}, $\Call{GET}{}$ and $\Call{PUT}{}$ (see Table~\ref{tab:atomic_subroutines}).
The reductions are identical in the MC$+$ and backup simulations: both are coordinated by the leader and differ only in their implementations of these subroutines.

\begin{table}[H]
\centering
\begin{tabular}{l||l}
\textbf{Subroutine} & \textbf{Description} \\\hline\hline
$\Call{PUT}{r,i,b}$ & sets the $i$-th bit of register $r$ to $b$ \\
$\Call{GET}{r,i}$ & returns the value of the $i$-th bit of register $r$\\
\end{tabular}
\caption{Atomic subroutines.}
\label{tab:atomic_subroutines}
\end{table}

\noindent\textbf{Atomic subroutines $\Call{PUT}{}$ and $\Call{GET}{}$.}
In MC$+$, each atomic access completes correctly whp in one round; otherwise, an anomaly is signalled (Protocols~\ref{alg:put} and~\ref{alg:get}). In the backup, the leader directly accesses the designated register-bit agent, guaranteeing correctness in $O(n)$ expected time (Protocols~\ref{alg:sput} and~\ref{alg:sget}). This modularisation leaves the higher-level instruction logic identical, with correctness following from that of the atomic accesses. Their detailed analyses appear in Subsections~\ref{ss:MC$+$_putget} and~\ref{ss:stable_putget}.

The leader maintains an instruction pointer $ip$ indicating the current instruction, and emulates the program step by step using the reductions below.
Unless $ip$ is explicitly modified by an instruction, the leader increments it by~$1$ after completing the instruction.
We first present the most involved reduction, that of the copy instructions: for indirect addressing, the leader must retrieve a register address from a register and store it in its state, which contributes a factor~$O(\rho'w)$ to the number of leader and working set agents states, whereas the remaining reductions, given below, require only $O(w)$ additional leader states.
\smallskip

\noindent\textbf{Indirect and direct copy} $\Call{COPY}{\cdot}$.
To implement both Cook--Reckhow variants of the copy instruction, we first
reduce them to two simpler subroutines: $\Call{SET}{r, rp}$ and the direct
$\Call{COPY}{rs, rd}$.
The former obtains the value stored in register $rp$ by successive
$\Call{GET}{rp, i}$ subroutines and stores it in the leader's state as the register index~$r$, while the latter performs a direct copy of the contents of register $rs$ into register $rd$.
The address bound is $a_{\max}=\rho$ for source instructions and $a_{\max}=\rho'$ for trusted input/output wrappers, as determined by the fixed control context. $\Call{SET}{}$ checks all $w$ bits and rejects an index outside $\{0,\ldots,a_{\max}-1\}$, including a negative word. An \textsc{InvalidAddress} outcome aborts the request: it signals an MC$+$ anomaly or suspends the current backup instance while its priority rank rules remain active. It cannot occur in a correct execution under Definition~\ref{P-programs}. The partial index is stored only below $a_{\max}$, so this check uses $O(\rho')$ index states.

\par
\begingroup\setlength{\intextsep}{0pt}%
\noindent
\begin{minipage}[t]{0.485\textwidth}
\begin{algorithm}[H]
\begin{algorithmic}[1]
\Require{$rp$: register; $r$: leader's index}
\Ensure{Valid $r=[rp]$, or \textsc{InvalidAddress}}
\Function{SET}{$r, rp$}
\State $r\gets 0$
\For{$i \gets 0$ \textbf{to} $w-1$}
\State{$b \gets\Call{GET}{rp, i}$}
\If{$b=1$ \textbf{and} $r+2^i\ge a_{\max}$}
    \State \Return \textsc{InvalidAddress}
\EndIf
\State{$r\gets r +2^i b$}
\EndFor
\EndFunction
\end{algorithmic}
\caption{$\Call{SET}{r, rp}$}\label{alg:set}
\end{algorithm}
\end{minipage}\hfill
\begin{minipage}[t]{0.485\textwidth}
\begin{algorithm}[H]
\begin{algorithmic}[1]
\Require{$rs, rd$: registers}
\Ensure{$[rd]=[rs]$}
\Function{COPY}{$rs, rd$}
\For{$i \gets 0$ \textbf{to} $w-1$}
\State{$b \gets\Call{GET}{rs, i}$}
\State{$\Call{PUT}{rd, i, b}$}
\EndFor
\EndFunction
\end{algorithmic}
\caption{$\Call{COPY}{rs, rd}$}\label{alg:copy}
\end{algorithm}
\end{minipage}
\endgroup
\par\medskip

Using these two building blocks, the instructions $\Call{COPY}{i, rp, d, rd}$
and $\Call{COPY}{d, rs, i, rp}$ (Protocols~\ref{alg:copy_id} and~\ref{alg:copy_di}, page~\pageref{alg:copy_id}) are decomposed into two parts: first obtaining
the indirect source or destination register index with $\Call{SET}{r, rp}$,
and then performing a direct copy.

Here we give the pseudocode of the reductions of $\Call{COPY}{i, rp, d, rd}$ and $\Call{COPY}{d, rs, i, rp}$.

\begin{algorithm}[H]
\begin{algorithmic}[1]
\Require{$rp, rd$ - pointers to registers, $i,d$ - indirect/direct flags}
\Ensure{Register $rd$ stores the contents of register $r=[rp]$}
\Function{COPY}{$i, rp, d, rd$}
\State $\Call{SET}{r, rp}$
\State $\Call{COPY}{r, rd}$
\EndFunction
\end{algorithmic}
\caption{$\Call{COPY}{i, rp, d, rd}$}\label{alg:copy_id}
\end{algorithm}

\begin{algorithm}[H]
\begin{algorithmic}[1]
\Require{$rs, rp$ - pointers to registers, $i,d$ - indirect/direct flags}
\Ensure{Register $r=[rp]$ stores the contents of register $rs$}
\Function{COPY}{$d, rs, i, rp$}
\State $\Call{SET}{r, rp}$
\State $\Call{COPY}{rs, r}$
\EndFunction
\end{algorithmic}
\caption{$\Call{COPY}{d, rs, i, rp}$}\label{alg:copy_di}
\end{algorithm}

\paragraph{Further instruction reductions.}
The following paragraphs give the remaining reductions of the RAM instructions to the atomic subroutines $\Call{PUT}{}$ and $\Call{GET}{}$.

The leader maintains an instruction pointer $ip$ indicating the instruction currently being executed, and emulates the program step by step using the reductions below.
Unless $ip$ is explicitly modified by an instruction, the leader increments it by~$1$ after completing the instruction.
\smallskip

\noindent\textbf{Conditional jump} $\Call{JGE}{r, Iz}$.
The conditional jump instruction is reduced to a single call to
$\Call{GET}{r,w-1}$, which retrieves the most significant bit of register $r$.
Since values are represented in U2 encoding, this bit determines the sign of
the stored value: if it is $0$, the value is non-negative. In this case, the
jump is taken and the leader updates the instruction pointer to the target
instruction $Iz$; otherwise, it proceeds to the next instruction $ip+1$.
The Cook-Reckhow model~\cite{CookR73} instead tests whether the register content is positive; each test is expressible by the other with $O(1)$ instructions.

\begin{algorithm}[H]
\begin{algorithmic}[1]
\Require{$r$ - pointer to a register, $Iz$ - program instruction index}
\Ensure{Jump to instruction $Iz$ if the content of $r$ is non-negative}
\Function{JGE}{$r, Iz$}
\State $b \gets \Call{GET}{r, w-1}$
\LineIf{$b=0$}{$ip\gets Iz$}
\EndFunction
\end{algorithmic}
\caption{$\Call{JGE}{r, Iz}$}\label{alg:jnz}
\end{algorithm}

\noindent\textbf{Loading a constant into a register} $\Call{LOAD}{C, rd}$.
Let $c_0, c_1, \dots,c_{w-1}$ be the complete $w$-bit U2 representation of
a predefined constant $C$, encoded in the transition function, ordered from
the least to the most significant bit.
The reduction for the $\Call{LOAD}{C, rd}$ instruction assigns the
value $C$ to register $rd$ bit by bit, where in each iteration $i$ the agent
representing the $i$-th bit of $rd$ adopts value $c_i$ via $\Call{PUT}{rd, i, c_i}$.
Constants independent of $w$ need only a fixed number of bits in the program code and the transition function; for larger $w$, they are extended by copies of the sign bit.

\begin{algorithm}[H]
\begin{algorithmic}[1]
\Require{$C=c_0\ldots c_{w-1}$ - a constant in U2, $rd$ - pointer to a register}
\Ensure{Register $rd$ contains the value $C$}
\Function{LOAD}{$C, rd$}
\For{$i \gets 0$ \textbf{to} $w-1$}
\State $\Call{PUT}{rd, i, c_i}$
\EndFor
\EndFunction
\end{algorithmic}
\caption{$\Call{LOAD}{C, rd}$}\label{alg:load}
\end{algorithm}

\noindent\textbf{Addition and subtraction} $\Call{ADD}{rs1, rs2, rd}$
\textbf{and} $\Call{SUB}{rs1, rs2, rd}$.
Both instructions are emulated bit by bit: at each position, the leader retrieves the two source bits using $\Call{GET}{}$, computes the result bit and the new carry locally, and
writes the result bit using $\Call{PUT}{}$. The final carry is discarded, so the operations are performed modulo~$2^w$; in particular, signed overflow wraps around, as is standard for U2 arithmetic.

\begin{algorithm}[H]
\begin{algorithmic}[1]
\Require{$rs1, rs2, rd$ - pointers to registers}
\Ensure{Register $rd$ stores the $w$-bit sum of the values of registers
$rs1, rs2$}
\Function{ADD}{$rs1, rs2, rd$}
\State $carry\gets 0$
\For{$i \gets 0$ \textbf{to} $w-1$}
\State{$b1 \gets\Call{GET}{rs1, i}$}
\State{$b2 \gets\Call{GET}{rs2, i}$}
\State{$b\gets (b1+b2+carry) \bmod 2$, \quad $carry \gets \lfloor(b1+b2+carry)/2\rfloor$}
\State{$\Call{PUT}{rd, i, b}$}
\EndFor
\EndFunction
\end{algorithmic}
\caption{$\Call{ADD}{rs1, rs2, rd}$}\label{alg:add}
\end{algorithm}

Subtraction uses the U2 identity $a-b=a+\neg b+1$.

\begin{algorithm}[H]
\begin{algorithmic}[1]
\Require{$rs1, rs2, rd$ - pointers to registers}
\Ensure{Register $rd$ stores the $w$-bit difference of the values of
registers $rs1, rs2$}
\Function{SUB}{$rs1, rs2, rd$}
\State $carry\gets 1$
\For{$i \gets 0$ \textbf{to} $w-1$}
\State{$b1 \gets \Call{GET}{rs1, i}$}
\State{$b2 \gets \Call{GET}{rs2, i}$}
\State{$b\gets (b1+(1-b2)+carry) \bmod 2$, \quad $carry \gets \lfloor(b1+(1-b2)+carry)/2\rfloor$}
\State{$\Call{PUT}{rd, i, b}$}
\EndFor
\EndFunction
\end{algorithmic}
\caption{$\Call{SUB}{rs1, rs2, rd}$}\label{alg:sub}
\end{algorithm}

The atomic subroutines and the RAM instructions are also used outside the emulation of the program itself: in the conversions and as macros in other subprotocols.
We also use the instructions $\Call{INC}{r}$ and $\Call{DEC}{r}$, which increase and decrease the content of register~$r$ by~$1$. They are emulated as $\Call{ADD}{}$ and $\Call{SUB}{}$ with the second operand replaced by the constant~$1$, encoded in the transition function, so each requires $2w$ atomic subroutines.
Table~\ref{tab:ram_reductions} and Fact~\ref{f:RAM_instructions} summarise the reductions.

\begin{table}[H]
\centering
\begin{tabular}{l||l||l}
\textbf{Instruction} & \textbf{Reduction} & \textbf{Atomic subroutines} \\\hline\hline
$\Call{JGE}{r,Iz}$ & read the sign bit & $1$ \\
$\Call{LOAD}{C,rd}$ & write the $w$ bits of $C$ & $w$ \\
$\Call{COPY}{i,rp,d,rd}$, $\Call{COPY}{d,rs,i,rp}$ & $\Call{SET}{}$, then direct copy & $\le 3w$ \\
$\Call{ADD}{rs1,rs2,rd}$, $\Call{SUB}{rs1,rs2,rd}$ & bitwise, with carry & $3w$ \\
$\Call{INC}{r}$, $\Call{DEC}{r}$ & bitwise, with carry & $2w$ \\
\end{tabular}
\caption{Reduction of RAM instructions to atomic subroutines.}
\label{tab:ram_reductions}
\end{table}

\begin{fact}\label{f:RAM_instructions}
Emulation of each RAM instruction (including $\Call{INC}{}$ and
$\Call{DEC}{}$) requires the execution of at most $3w$ atomic subroutines
$\Call{GET}{ }$ and $\Call{PUT}{ }$.
\end{fact}

\providecommand{\newtxt}[1]{\textcolor{teal}{#1}}
\providecommand{\oldtxt}[1]{\textcolor{orange!85!black}{#1}}
\providecommand{\clnote}[1]{\textcolor{violet}{[#1]}}

\section{Efficient \texorpdfstring{MC$+$}{MC+} Simulation}\label{s:MC$+$_sim}\label{app:MC$+$_sim}

\subsection{Overview}\label{ss:MC$+$_overview}

The efficient simulation is an MC$+$ protocol~\cite{gasieniec_et_al:LIPIcs.DISC.2026.17}: whp it is correct and terminates in $\poly(w)$ time; otherwise, it signals a protocol or clock \emph{anomaly}. 
Two results of~\cite{gasieniec_et_al:LIPIcs.DISC.2026.17} reduce the proof of Theorem~\ref{t:main} to building such a protocol from polylogarithmically many MC$+$ stages, following the scheme of Figure~\ref{fig:sim_scheme} and Protocol~\ref{alg:MC$+$_simulation_scheme}, and combining it with a silent backup.

\begin{restatable}{theorem}{tmcframe}[(i) Theorem~3 and (ii) Theorem~2 in~\cite{gasieniec_et_al:LIPIcs.DISC.2026.17}]\label{th:MCp_framework}
\leavevmode
\begin{enumerate}[label=(\roman*), ref=\thetheorem(\roman*), nosep]
\item\label{th:combine_MC$+$} The sequential composition of polylogarithmically many MC$+$ protocols is an MC$+$ protocol.
\item\label{th:stable+MC$+$} An efficient MC$+$ protocol and a polynomial-time stable protocol for the same problem can be combined into a protocol that is both efficient and stable; if the stable protocol is silent, so is the combined one.
\end{enumerate}
\end{restatable}

Our MC$+$ protocol modifies the clock of~\cite{gasieniec_et_al:LIPIcs.DISC.2026.17} and ends with a terminal verification round, performing all pending anomaly checks before the clock stops. The remainder of this section shows that it retains the properties required by Theorem~\ref{th:stable+MC$+$}: every malfunction is eventually signalled, the expected time to completion or detection is polynomial from every reachable configuration, and a correct execution becomes silent.
We prove the following theorem, and all other claims of this section, below.

\begin{restatable}{theorem}{tmcp}\label{t:MC$+$sim}
For every $\mathcal{P}$-program, there exists an efficient MC$+$ population protocol that, whp, correctly simulates the program in
$\tau=O(w\,(T_Rw+T_{\mathrm{RAM}}+T_P))$ rounds of $\Theta(w)$ time each, plus one register initialisation round lasting $O(\rho'w)$ time.
The protocol uses $O(\rho'w\tau)=O(\rho'w^2\,(T_Rw+T_{\mathrm{RAM}}+T_P))$ states.
\end{restatable}

Besides leader election, adopted from~\cite{gasieniec_et_al:LIPIcs.DISC.2026.17}, the MC$+$ simulation rests on four techniques.

\noindent\textbf{(1) Bit access in one round.}
By Subsection~\ref{ss:ram_instructions}, every RAM instruction reduces to at most $3w$ calls of $\Call{PUT}{}$ and $\Call{GET}{}$, so it suffices to make a single bit access fast and anomaly-detecting.
In the MC$+$ variants $\Call{PUT\textsuperscript{e}}{r,i,b}$ and $\Call{GET\textsuperscript{e}}{r,i}$, the leader opens a round with a request that spreads through the working set by an epidemic; the register agent $(R;r,i,\cdot)$, which holds the $i$-th bit of register~$r$, applies the update or, for $\Call{GET\textsuperscript{e}}{}$, starts a second epidemic that carries the bit back to the leader.
Since the working set is a constant fraction of the population (Fact~\ref{f:free_agents}), both complete whp within one round; a register agent that has not processed a $\Call{PUT\textsuperscript{e}}{}$ request, or a leader that has not received the requested bit, signals an anomaly (Lemma~\ref{l:MC$+$_getput}).
Together with the address checks, these subroutines ensure correct instruction emulation in the absence of an anomaly, using at most $3w$ rounds per instruction (Lemma~\ref{l:MC$+$_ram_instructions}).

\noindent\textbf{(2) A cheaper clock.}
In~\cite{gasieniec_et_al:LIPIcs.DISC.2026.17}, rounds advance by counting leader interactions, which multiplies the leader's states by $\Theta(\log n)$ per round.
We count interactions only during MC$+$ leader election ($\Theta(w)$ rounds, $O(w^2)$ states; Theorem~\ref{t:MC$+$_le}).
Register initialisation, whose duration $O(\rho'w)$ may exceed $\Theta(\log n)$, then occupies a single \emph{long} round, after which the protocol switches to the improved clock based on the phase-clock mechanism of~\cite{DBLP:journals/dc/AngluinAE08a} with a constant number of phases.
Each round then costs $O(1)$ states and lasts $\Theta(w)$ time, and whp contains an interval of $\Omega(\log n)$ time in which all agents share the same round number (Theorem~\ref{t:clock}), which is what every single-round subprotocol relies on.

\noindent\textbf{(3) Weights in one round.}
Conversions use the \emph{system of weights}: disjoint sets of agents $P_0,\ldots,P_{w-2}$ with $|P_k|=2^k$, whose unions represent every value below $2^{w-1}$.
Since $w$ is known, we replace the sequential $O(\log n)$-round construction of~\cite{gasieniec_et_al:LIPIcs.DISC.2026.17} by a parallel single-round one: a delegating agent hands mass $2^k$ of $P_k$ to a fresh agent for $k=w-2,\ldots,1$, and every mass is halved with free agents until each agent holds unit mass.
The process may be represented as a binary tree of depth at most $w$ with fewer than $n$ leaves, so a Chernoff and union bound give completion whp in one round, with $\Theta(w^2)$ states (Lemma~\ref{l:weights_MC$+$}).

\noindent\textbf{(4) Conversion by binary search.}
Extending the parity computation of~\cite{gasieniec_et_al:LIPIcs.DISC.2026.17} from one bit to retrieving all $w$ bits, $\Call{U2Bconversion\textsuperscript{e}}{i}$ determines $|X_i|$ bit by bit from the most significant one: it adds $P_\ell$ to an auxiliary subpopulation $Z$ and compares $|Z|$ with $|X_i|$ by the MC$+$ majority protocol (Theorem~\ref{t:MC$+$_majority}), keeping $P_\ell$ and writing bit~$1$ if $|Z|\le|X_i|$, and otherwise removing it and writing~$0$; this takes $O(w^2)$ rounds per input variable (Lemma~\ref{l:MC$+$_input_conversion}).
The invariant is $0\le |X_i|-|Z|<2^{\ell+1}$ before bit~$\ell$, initially $\ell=w-2$ and $Z=\emptyset$. The comparison and choice reduce the residual to $[0,2^\ell)$; after bit~$0$ it is zero. The untouched sign bit remains~$0$ from register initialisation, so the stored word equals~$|X_i|$.
Conversely, each weight acts as a stamp: for each bit $\ell$ of $r_i^{out}$ equal to~$1$, every agent of $P_\ell$ first adopts an additional component $out_i$ and then passes it to a free agent, so all $2^\ell$ copies are stamped in parallel and $2^\ell$ output agents are created in one round, while $P_\ell$ is preserved for later use (Lemmas~\ref{l:MC$+$_out} and~\ref{l:MC$+$output_conv}).

Summing over the stages gives $\tau$ rounds, dominated by conversions and program emulation.
The state bound $O(\rho'w\tau)$ comes from agents that must record a register index and a bit index in every round: the register agents, and the leader and working set while emulating RAM instructions with indirect addressing.
Our technical refinement reduces this bound by a factor $\Theta(\min(\rho',w))$.

\begin{restatable}{theorem}{tmcpr}\label{t:MC$+$sim_refined}
The MC$+$ simulation of Theorem~\ref{t:MC$+$sim} can be implemented in the same number of rounds and with $O\!\left((\rho'+w)\,\tau\right)$ states.
\end{restatable}

The key idea is a \emph{fixed schedule}.
We pad every instruction emulation and every output conversion to exactly $3w$ rounds, and every iteration of the input conversion to a fixed length; we call these segments \emph{passes}. The instruction tag remains fixed throughout its pass, including padding; the updated instruction pointer takes effect in the next pass.
The bit index of every atomic subroutine call is then a function of the round, so the leader and the working set may store only a register index, i.e., $O(\rho')$ states per round.
Register agents, whose identities $(r,i,b)$ alone take $\Theta(\rho'w)$ states, instead count \emph{coarse-grained rounds}, one per pass of at least $w$ rounds, and keep an $O(1)$-size status locating the current round relative to the at most three rounds of the pass in which their bit can be accessed.
They update it from their partners' exact rounds, checking pending obligations and skipped active rounds before each update and on termination, so failure detection is preserved.
Since there are only $O(\tau/w)$ passes, register agents use $O(\rho'\tau)$ states.
Combining Theorem~\ref{t:MC$+$sim_refined} with the backup of Section~\ref{s:stable_sim} via Theorem~\ref{th:stable+MC$+$} improves the state bound of Theorem~\ref{t:main} accordingly.

\subsection{Simulation Scheme}\label{ss:MC$+$_scheme}
The following subsections give the full description and proofs of the MC$+$ simulation outlined in Subsection~\ref{ss:MC$+$_overview}.
\bigskip

The fast simulation is an MC$+$ protocol~\cite{gasieniec_et_al:LIPIcs.DISC.2026.17} composed of MC$+$ subprotocols that implement successive operations of the simulated $\mathcal{P}$-program.
It follows the scheme presented in Protocol~\ref{alg:MC$+$_simulation_scheme}.
First, the MC$+$ leader election protocol is executed, after which the elected leader initialises all required registers (Subsection~\ref{ss:MC$+$_clock_le}).
Next, the leader initiates the MC$+$ weight creation protocol, which constructs disjoint subsets of agents of sizes $2^0,2^1,\ldots,2^{w-2}$, called \emph{weights} (Subsection~\ref{ss:MC$+$_clock_le}).
Together with the MC$+$ majority protocol (Subsection~\ref{ss:MC$+$_clock_le}), these weights enable the efficient conversion of the input values from unary to binary, with the resulting values stored bit by bit in the input array registers (Subsection~\ref{ss:MC$+$_inout}).
The RAM instructions are then emulated according to the reductions described in Subsection~\ref{ss:ram_instructions}, using the MC$+$ variants of the atomic subroutines $\Call{GET}{}$ and $\Call{PUT}{}$ (Subsection~\ref{ss:MC$+$_putget}).
Finally, the values stored in the output array are converted from binary to unary using the weights, yielding the output of the simulation and the program (Subsection~\ref{ss:MC$+$_inout}).

The MC$+$ simulation produces the correct output whp in $\poly(w)=\poly\log n$ time and uses $\poly(w)$ states, with both bounds depending on the running time and space usage of the simulated $\mathcal{P}$-program.
With negligible probability, an anomaly may occur; when it is detected, the protocol switches to the slower but stable backup simulation described in Section~\ref{s:stable_sim}.

The fast execution is bounded even on faulty histories. We pad the modified program's execution to
\[
T^*=T_{\mathrm{RAM}}+2(T_R+T_P)+2
\]
instruction slots, using $3w$ rounds per slot and idle rounds after the program halts. A program that has not halted by the last slot signals an anomaly. Input-conversion branches and output conversions are also padded to fixed lengths, as detailed in the refined state analysis. Thus the round schedule has a prescribed finite endpoint determined by the parameters, and an instruction-slot index is determined by the round rather than stored separately.
After the last output round, the leader initiates a terminal verification round. Every agent performs all pending end-of-round and skipped-round checks before adopting the terminal state. The terminal marker continues to propagate, but the clock no longer advances; outstanding anomaly witnesses and the rule detecting two leaders are retained. This adds only a constant number of rounds and ensures that stopping the computation cannot suppress an anomaly.

\begin{algorithm}[H]
\begin{algorithmic}[1]
\Require{A modified $\mathcal{P}$-program; $T=T_R+T_{\mathrm{RAM}}+T_P$, $\rho'$, $w$; a leaderless population of $n$ agents}
\Ensure{Correct simulation of the given program whp, or an eventual \textsc{Anomaly} signal otherwise}
\Function{MC$+$ Simulation}{}
\State $\Call{LeaderElection}{}$
\State $\Call{RegInit}{\rho',w}$
\State $\Call{WeightCreation}{w}$
\For{$i\gets 1$ \textbf{to} $T_R$}
\State $\Call{U2Bconversion\textsuperscript{e}}{i}$ \Comment{unary-to-binary conversion of $i$-th input}
\EndFor
\State $ip\gets 0$
\For{each of the $T^*$ prescribed instruction slots}
\If{$ip\neq\mathsf{END}$}
\State Emulate the RAM instruction indicated by $ip$ and update $ip$.
\EndIf
\State Pad the slot with idle rounds to a total of $3w$ rounds.
\EndFor
\LineIf{$ip\neq\mathsf{END}$}{\textbf{signal} \textsc{Anomaly}}
\For{$i\gets 1$ \textbf{to} $T_P$}
\State $\Call{B2Uconversion\textsuperscript{e}}{i}$ \Comment{binary-to-unary conversion of $i$-th output}
\EndFor
\State Enter the terminal verification round and stop advancing the clock.
\EndFunction
\end{algorithmic}
\caption{$\Call{MC+ Simulation}{}$ Protocol}\label{alg:MC$+$_simulation_scheme}
\end{algorithm}

In the remainder of this section, we develop the techniques needed to establish the following theorem on efficient MC$+$ simulation.
The proof of the theorem is provided in Subsection~\ref{ss:MC$+$summary}.

\tmcp*

\subsection{\texorpdfstring{MC$+$}{MC+} Protocols: Overview and Refinements}\label{ss:MC$+$_clock_le}

MC$+$ protocols were introduced in~\cite{gasieniec_et_al:LIPIcs.DISC.2026.17}.
They are fast Monte Carlo population protocols that execute correctly whp in $\poly(\log n)$ time and additionally guarantee that any malfunction is eventually signalled by an \emph{anomaly}.
MC$+$ protocols facilitate the modular design of complex computations in population protocols, as compositions of MC$+$ protocols are themselves MC$+$ protocols.
Each subprotocol uses a \emph{clocking mechanism} to track a prescribed polylogarithmic number of consecutive computation rounds, after which it becomes silent, provided that no anomaly occurs.
This allows subprotocols to be executed sequentially, with each building on the results of its predecessors.
If all subprotocols execute correctly, their composition also produces the correct result.
If a subprotocol malfunctions, an event of negligible probability, detection of the resulting anomaly triggers a switch to the stable backup.
We distinguish two types of anomalies:
\begin{enumerate}
    \item a \emph{protocol anomaly}, arising from a failure of the subprotocol's own computation;
    \item a \emph{clock anomaly}, detected when interacting agents have round numbers that differ by more than one, or when a post-election leader encounters a fine-clock follower with a later timestamp.
\end{enumerate}

\noindent\textbf{Clock}
An essential component of an MC$+$ protocol is a \emph{clock} that organises the computation into successive \emph{rounds}.
Each round lasts $\Theta(\log n)=\Theta(w)$ time whp.
In the original MC$+$ construction~\cite{gasieniec_et_al:LIPIcs.DISC.2026.17}, rounds advance by counting leader interactions, which increases the number of leader states by a factor of $\Theta(\log n)$.

We propose to employ a refined clock.
In our construction, interaction counting is used to advance rounds only during leader election.
Immediately afterwards, the registers are initialised in a single \emph{long} round (see page~\pageref{reg_init_paragraph}).
Once initialisation is complete, the protocol switches to a clock based on the phase-clock mechanism of~\cite{DBLP:journals/dc/AngluinAE08a}.
By Theorem~1 of~\cite{DBLP:journals/dc/AngluinAE08a}, this mechanism can provide intervals of length $\Theta(\log n)$ whp, with an arbitrarily large constant lower bound on their duration.
We implement the clock using a constant number $m$ of phases.
Each agent has a clock component of the form $(L|F,t,x)$, where $L$ and $F$ denote a leader and a follower, respectively, $t$ is the current round, and $x\in\{1,\ldots,m\}$ is the current phase.

The clock transitions are as follows, where $\star$ denotes either $L$ or $F$:
\begin{align*}
    &(\star,t,x)+(F,t,y)\rightarrow (\star,t,x)+(F,t,x), &\text{if } y<x,\\
    &(\star,t+1,x)+(F,t,y)\rightarrow (\star,t+1,x)+(F,t+1,x),\\
    &(L,t,x)+(F,t,x)\rightarrow (L,t,x+1)+(F,t,x), &\text{if } x<m,\\
    &(L,t,m)+(F,t,m)\rightarrow (L,t+1,1)+(F,t,m),\\
    &(\star,t',x)+(\star,t,y)\rightarrow \text{clock \textsc{Anomaly}}, &\text{if } |t'-t|>1.
\end{align*}

After leader election, the following additional check has priority in both interaction orientations: if a leader meets a fine-clock follower whose timestamp $(t',y)$ is lexicographically later than the leader's $(t,x)$, it signals a clock anomaly. In a correctly initialised unique-leader execution, the leader creates every new timestamp, so this check never fires. It uses only the existing clock fields and prevents an inherited clock discrepancy from leaving the unique leader unable to advance.

Combining Theorem~1 of~\cite{DBLP:journals/dc/AngluinAE08a} with its epidemic propagation bound yields the following guarantee.

\begin{theorem}\label{t:clock}
For any fixed constants $a,c,\eta>0$, there exist constants $m$ and $D$, depending only on $a,c,\eta$, such that the following holds for all sufficiently large $n$.
Starting from the standard phase-clock initialisation with a unique leader, with probability at least $1-n^{-\eta}$, each of the first $ n^a$ rounds lasts at most $D\log n$ time and contains an interval of at least $c\log n$ time during which all agents have the same round number.
\end{theorem}

We also need a bound outside the high-probability event. With a unique leader, an ahead-of-leader follower triggers the additional check in expected time $O(n)$. Otherwise, any phase advance is enabled after at most two specified leader--follower interactions: the first brings that follower to the leader's phase, and the second advances the leader. Each such interaction has expected waiting time $O(n)$. If an intervening discrepancy signals an anomaly, the fast execution stops earlier. Since the number of phases per round is constant and the round budget is finite, completing the clock or detecting an anomaly has a polynomial expected residual time, uniformly over reachable configurations. Every leader that starts register allocation retains its post-election leader flag until fallback. A subsequent interaction with any other leader or election candidate, including one still completing election, signals an anomaly. The expected time until such an anomaly is detected or no other candidate remains is $O(n)$. Once no other candidate remains, only one allocation has created auxiliary roles, so Fact~\ref{f:free_agents} applies and the remaining allocations take expected time $O(\rho'w)$ by Lemma~\ref{l:reg_init}. Thus register allocation also has a uniform polynomial expected residual bound. Terminal propagation executes pending checks before their evidence can be overwritten.

Consequently, if $A=O(\rho'w+\tau\log n)$ is a high-probability bound for successful completion of the fast simulation and $D(n)=n^{O(1)}$ bounds the expected remaining detection, switching, and backup time from any reachable configuration, then
\[
\mathbb{E}[\text{stabilisation time}]
\le A+n^{-\eta}D(n)=O(A),
\]
by choosing the failure exponent $\eta$ larger than the fixed degree in $D$. This is the expected-time accounting required by the MC$+$ combination theorem below; eventual detection alone would not suffice.

The following theorems, formulated and proved in~\cite{gasieniec_et_al:LIPIcs.DISC.2026.17}, form the foundation of the MC$+$ simulation.
Theorem~\ref{th:stable+MC$+$} establishes that an efficient MC$+$ protocol can be combined with a stable backup to obtain an efficient and stable population protocol.
Theorem~\ref{th:combine_MC$+$} establishes that MC$+$ protocols can be composed to create more complex protocols that are themselves MC$+$.

\tmcframe*

We next describe the MC$+$ protocols from~\cite{gasieniec_et_al:LIPIcs.DISC.2026.17} that we adapt for the simulation.
The leader election and majority protocols require only minor modifications, whereas our weight creation protocol is substantially simpler and completes in fewer rounds than the original construction.
\smallskip

\noindent\textbf{MC$+$ Leader Election}
The simulation begins with the MC$+$ leader election protocol of~\cite{gasieniec_et_al:LIPIcs.DISC.2026.17}, which is based on a scheme used in several prior works~\cite{DBLP:journals/dc/AngluinAE08a, DBLP:journals/jacm/GasieniecS21, DBLP:conf/stoc/BerenbrinkGK20}.
It operates over $\Theta(\log n)$ rounds and is designed so that all leaders cannot disappear, as this would cause the clock to stop.
In our simulation, the number of rounds is set to a sufficiently large constant multiple of $w$.
In successive rounds, the potential leaders perform coin tosses using randomness supplied by the scheduler.
If at least one potential leader obtains heads in a round, those obtaining tails become followers.
Each outcome is associated with its round number, ensuring that at least one leader persists.
Although more efficient leader election protocols exist, adapting one to the MC$+$ variant would not improve the overall asymptotic complexity of our simulation.

With negligible probability, more than one leader remains after the prescribed rounds.
Since subsequent MC$+$ subprotocols assume a unique leader, this may lead to an incorrect computation.
To detect this case, any subsequent interaction between two leaders signals an anomaly.
Such an interaction occurs eventually with probability~$1$, and the expected time until it occurs is $O(n)$.

\begin{theorem}\label{t:MC$+$_le}
There exists an MC$+$ leader election protocol that elects a unique leader whp in $\Theta(\log n)=\Theta(w)$ rounds and uses $O(\log^2 n)=O(w^2)$ states.
\end{theorem}

\begin{proof}
Correctness and round complexity follow from Theorem~4 of~\cite{gasieniec_et_al:LIPIcs.DISC.2026.17} and the relation $w=\Theta(\log n)$.
The state bound accounts for the clock that counts leader interactions.
This clock requires $\Theta(\log n)=\Theta(w)$ states per round for the leader.
\end{proof}

\noindent\textbf{MC$+$ Weight Creation}
We now describe the MC$+$ protocol for creating the \emph{system of weights}: pairwise disjoint sets of agents $P_0,P_1,\ldots,P_{w-2}$, where $|P_i|=2^i$.
Together, these sets contain $2^{w-1}-1$ agents, allowing unions of weights to represent every integer from $0$ to $2^{w-1}-1$, that is, every non-negative $w$-bit value.
The weights support both the conversion of unary input values to binary representations in registers and the final conversion of binary output values back to unary.

The word length $w$, and hence the required number of weights, is known in advance.
This allows us to propose a simpler and faster construction than the one in~\cite{gasieniec_et_al:LIPIcs.DISC.2026.17}.
Whereas that protocol creates weights sequentially over $O(\log n)$ rounds, ours creates all $w-1$ weights in a single round.

The protocol $\Call{WeightCreation}{w}$ is initiated by the leader.
In its first interaction with a free agent, the leader delegates the creation of all weights by assigning that agent state $p^*(w-2)$.
The subsequent weight creation process is governed by the following transitions:
\begin{align}
&p^*(k)+\varphi\rightarrow p^*(k-1)+p_k(k), &&\text{for } k>1,\label{eq:MC$+$_weight_delegate}\\
&p^*(1)+\varphi\rightarrow p_0(0)+p_1(1),\label{eq:MC$+$_weight_base}\\
&p_k(i)+\varphi\rightarrow p_k(i-1)+p_k(i-1), &&\text{for } i>0.\label{eq:MC$+$_weight_split}
\end{align}

We say that an agent in state $p_k(i)$ \emph{holds mass} $2^i$ belonging to weight $P_k$.
The total mass of $P_k$ is $2^k$, initially held by a single agent in state $p_k(k)$.
This mass is repeatedly split until it is distributed among $2^k$ agents, each holding unit mass.

The agent in state $p^*(k)$ is responsible for initiating the creation of $P_k,P_{k-1},\ldots,P_0$.
When $k>1$, an interaction with a free agent assigns the entire mass $2^k$ of $P_k$ to that agent, which adopts state $p_k(k)$, according to rule~\eqref{eq:MC$+$_weight_delegate}.
At the same time the delegating agent adopts state $p^*(k-1)$ and continues with the remaining weights.
Rule~\eqref{eq:MC$+$_weight_base} completes this delegation process by initiating $P_1$ and creating $P_0$.
Thus, the creation of weights is initiated while the previously assigned masses are being split.
Within each weight $P_k$, splitting proceeds according to rule~\eqref{eq:MC$+$_weight_split}.
An agent in state $p_k(i)$ with $i>0$ divides its mass equally with a free agent, leaving both in state $p_k(i-1)$.
The process ends when all agents in $P_k$ hold unit mass, that is, when they are in state $p_k(0)$.

All $w-1$ weights are created whp in one round.
Labelling the created weights requires $w = \Theta(\log n)$ states.
With negligible probability, the protocol signals an anomaly if, upon entering the next round, the leader failed to delegate the creation of weights or some follower remains in state $p_k(i)$ or $p^*(i)$, for $i>0$ and $k\in \{1, \ldots, w-2\}$.
In the absence of an anomaly, the result of the protocol is the family of sets $P_k$ for $k=0,\ldots, w-2$, where $P_k$ consists of agents in state $p_k(0)$.

\begin{lemma}\label{l:weights_MC$+$}
The protocol $\Call{WeightCreation}{w}$ correctly creates weights $P_0,P_1,\ldots,P_{w-2}$ whp in one round and signals an \textsc{Anomaly} otherwise.
It uses $\Theta(w^2)$ states.
\end{lemma}

\begin{proof}
We follow the approach of~\cite{gasieniec_et_al:LIPIcs.DISC.2026.17}.
By the choice of population size, at least $(1-1/C)n$ agents remain free throughout the construction.
Consequently, conditional on a specified active agent being selected as initiator, the probability that its partner is free is at least $1-1/C$.

After the leader's initial delegation, the construction can be represented by a binary tree rooted at $p^*(w-2)$.
A delegation step branches into the construction of the current weight and the delegation of the remaining weights; a splitting step branches into two equal masses belonging to the same weight.
The tree has $\sum_{k=0}^{w-2}2^k=2^{w-1}-1$ leaves, corresponding to the final unit-mass agents, and every root-to-leaf path has at most $w-1$ edges.
Including the leader's initial delegation adds at most one transition to each path.

Fix a path that is not yet complete.
Its next transition occurs when the agent responsible for that transition has an interaction with a free agent.
Conditional on the execution history, this happens in an interaction with probability at least $(1-1/C)/n$.
Hence, over $cn\log n$ interactions, progress along the path is bounded below by a binomial process of mean $(1-1/C)c\log n$, until the path is complete.
Since $w=\Theta(\log n)$, a Chernoff bound shows that the probability of completing fewer than the required $w$ transitions is at most $n^{-\eta-1}$ for a sufficiently large constant $c$.
A union bound over the fewer than $n$ root-to-leaf paths shows that all weights are created whp in $O(\log n)$ time.
If the construction does not finish, an unfinished delegation or splitting state remains and triggers an anomaly.
Finally, the states $p^*(k)$ and $p_k(i)$, with $0\le i\le k<w$, require $\Theta(w^2)$ states.
\end{proof}

\noindent\textbf{MC$+$ Majority}
Let $A$ and $B$ be two subpopulations of a population of $n$ agents.
The majority problem is to determine whether $|A|>|B|$, $|A|<|B|$, or $|A|=|B|$.
The MC$+$ majority protocol~\cite{gasieniec_et_al:LIPIcs.DISC.2026.17} employed in this work uses the standard operations of \emph{cancelling}, in which two agents with opposite biases become unbiased, and \emph{splitting}, in which a biased agent and an unbiased agent become two agents with the same bias.
These operations double the remaining bias over successive phases.
The protocol runs for $\Theta(\log n)$ rounds, producing the correct answer whp and signalling an anomaly otherwise.
In our simulation, the number of rounds is set to a sufficiently large constant multiple of $w$.
Apart from the clock, the protocol requires only $O(1)$ additional states per agent.

We believe that the best-known majority protocol~\cite{DBLP:conf/focs/DotyEGSUS21}, with $O(\log n)$ expected stabilisation time, could be adapted to the MC$+$ setting so that it operates in a constant number of rounds.
Such an adaptation would improve the complexity of our construction by reducing the time required for input conversion by a logarithmic factor.
We leave the development of such a protocol as an open problem, as determining whether such an adaptation is feasible, and constructing it if so, would introduce significant technical complications into the presentation of this work.

\begin{theorem}\label{t:MC$+$_majority}
There exists an MC$+$ majority protocol that executes correctly whp in $\Theta(\log n)=\Theta(w)$ rounds and uses $\Theta(\log n)=\Theta(w)$ states.
\end{theorem}

\begin{proof}
Correctness and round complexity follow from Theorem~6 of~\cite{gasieniec_et_al:LIPIcs.DISC.2026.17} and the relation $w=\Theta(\log n)$.
The computation requires only $O(1)$ additional states per round~\cite{gasieniec_et_al:LIPIcs.DISC.2026.17}.
At this stage, the clock uses $O(1)$ states per round, so tracking the $\Theta(w)$ rounds requires $\Theta(w)$ states in total.
\end{proof}

\noindent\textbf{Register Initialisation}\label{reg_init_paragraph}
The leader participates in an average of two interactions per unit of time.
By Lemma~\ref{l:reg_init}, initialising $\rho'=\poly(\log n)$ registers takes $O(\rho'w)$ time both in expectation and whp, and succeeds with probability~$1$.
We therefore allocate a single \emph{long} round immediately after leader election for this process.
Unlike an ordinary round, its duration is determined by completion of $\Call{RegInit}{}$ and may exceed $\Theta(\log n)$.
During register initialisation, the improved-clock components keep their standard initial values; once allocation is complete, the leader increments the round number by~$1$ and activates this clock, leaving the allocated registers unchanged, and the simulation continues with rounds of logarithmic duration.

\subsection{\texorpdfstring{MC$+$}{MC+} Emulation of RAM Instructions}\label{ss:MC$+$_putget}

In this subsection, we introduce efficient MC$+$ variants, $\Call{PUT\textsuperscript{e}}{}$ and $\Call{GET\textsuperscript{e}}{}$, of the atomic subroutines $\Call{PUT}{}$ and $\Call{GET}{}$.
These subroutines allow every RAM instruction to be emulated by an MC$+$ protocol via reductions of Subsection~\ref{ss:ram_instructions}.
We assume that a leader has been elected and that all required registers have been initialised.
The request and response states described below are additional state components; agents retain their register assignments and other persistent information.

The $\Call{PUT\textsuperscript{e}}{rd, i, b}$ subroutine (Protocol~\ref{alg:put}) completes successfully whp within one round.
At the beginning of the round, the leader adopts a state component indicating that the $i$-th bit of register $rd$ is to be set to $b$.
Every free agent in this round subsequently adopts the same component via epidemic process.
When a register agent enters the round, it adopts the state component $(put)$.
This component is cleared once the agent determines whether it holds the bit to be updated and, if so, performs the required change. If a register agent has not determined by the end of the round whether it is responsible for updating the bit, it signals an {\sc Anomaly}.

\begin{algorithm}[H]
\begin{algorithmic}[1]
\Require{$rd$ - register pointer, $i$ - bit index, $b\in\{0,1\}$}
\Ensure{The $i$-th bit of register $rd$ is set to $b$}
\Function{PUT\textsuperscript{e}}{$rd, i, b$}
\State The leader adopts component $(put,rd,i,b)$ at the start of round $t$.
\State \textbf{upon} entering round $t$ a follower adopts component $(put)$,
\LineIf{both agents are in round $t$}{
\State \hspace{\algorithmicindent} $(\varphi;put)+(\varphi|L;put,rd,i,b)\rightarrow (\varphi;put,rd,i,b)+(\varphi|L;put,rd,i,b)$,

\State \hspace{\algorithmicindent} $(R;put,rd,i,b')+(\varphi;put,rd,i,b)\rightarrow(R;rd,i,b)+(\varphi;put,rd,i,b).$
}
\LineIf{both agents are in round $t$ \textbf{and} $(rd',i')\neq (rd,i)$}{
\State \hspace{\algorithmicindent} $(R;put,rd',i',b')+(\varphi;put,rd,i,b)\rightarrow(R;rd',i',b')+(\varphi;put,rd,i,b)$. 
}
\LineIf{a register agent still holds component $(put)$ at the end of round $t$}{\State \hspace{\algorithmicindent} \textbf{signal} \textsc{Anomaly}}

\EndFunction
\end{algorithmic}
\caption{$\Call{PUT\textsuperscript{e}}{rd, i, b}$}\label{alg:put}
\end{algorithm}

The $\Call{GET\textsuperscript{e}}{rs, i}$ subroutine (Protocol~\ref{alg:get}) also completes successfully whp within one round.
At the beginning of the round, the leader adopts the component $(get, rs,i,?)$, requesting the value of the $i$-th bit of register $rs$, and propagates this request among the working set agents via an epidemic process.
When a working set agent interacts with the agent storing the requested bit, it initiates a second epidemic process that propagates the bit value to the other agents, eventually to the leader, during the remainder of the round.
The leader signals an {\sc Anomaly} if it does not receive the requested bit by the end of the round.

\begin{algorithm}[H]
\begin{algorithmic}[1]
\Require{$rs$ - register pointer, $i$ - bit index}
\Ensure{The leader learns the $i$-th bit of register $rs$}
\Function{GET\textsuperscript{e}}{$rs, i$}
\State The leader adopts component $(get,rs,i,?)$ at the start of round $t$.
\State \textbf{upon} entering round $t$ a follower adopts component $(get)$.
\LineIf{both agents are in round $t$}{
\State \hspace{\algorithmicindent} $(\varphi;get)+(\varphi|L;get,rs,i,z)\rightarrow (\varphi;get,rs,i,z)+(\varphi|L;get,rs,i,z)$,

\State \hspace{\algorithmicindent} $(R|\varphi;get,rs,i,b)+(\varphi|L;get,rs,i,?)\rightarrow(R|\varphi;get,rs,i,b)+(\varphi|L;get,rs,i,b).$}
\LineIfElse{the leader still holds $(get,rs,i,?)$ at the end of round $t$}{\State \hspace{\algorithmicindent} \textbf{signal} \textsc{Anomaly}}
{\textbf{return} $b$}
\EndFunction
\end{algorithmic}
\caption{$\Call{GET\textsuperscript{e}}{rs, i}$}\label{alg:get}
\end{algorithm}

Both MC$+$ subroutines may signal an anomaly with negligible probability.
Together with the address validation in $\Call{SET}{}$, their checks guarantee correct instruction emulation in the absence of an anomaly.
Their behaviour is summarised by the following lemma.

\begin{lemma}\label{l:MC$+$_getput}
The MC$+$ subprotocols $\Call{PUT\textsuperscript{e}}{}$ and $\Call{GET\textsuperscript{e}}{}$ complete correctly whp in one round; otherwise, an \textsc{Anomaly} is signalled.
\end{lemma}
\begin{proof}
At the beginning of the round of the $\Call{PUT\textsuperscript{e}}{rd,i,b}$ subroutine, the leader adopts the request component $(put,rd,i,b)$.
Each subsequent agent entering this round adopts the $(put)$ component.
The leader broadcasts the request $(put,rd,i,b)$ to all free agents through an epidemic process.
Since the free agents form a constant fraction of all agents, this epidemic reaches every one of them within $O(\log n)$ time whp.
The register agents then verify within $O(\log n)$ time whp whether they hold the bit to be updated and, if so, set it to $b$.
If a register agent fails to verify whether it is responsible for the update by the end of the round, it signals an {\sc Anomaly}.
Thus, in the absence of such an anomaly or clock anomaly, the requested bit is set correctly.

The $\Call{GET\textsuperscript{e}}{rs, i}$ protocol begins with the leader adopting the request component $(get, rs,i,?)$ at the beginning of round~$t$.
Every follower entering round~$t$ subsequently adopts the $(get)$ component.
The leader broadcasts the request $(get,rs,i,?)$ to all free agents through an epidemic in time $O(\log n)$ whp.
Consequently, the agent holding the $i$-th bit of the register $rs$ also receives this request during round~$t$ whp by an interaction with a free agent.
During the remainder of round~$t$, the component $(get,rs,i,b)$ spreads to all working set agents and the leader whp via a second epidemic process, also within time $O(\log n)$ whp.
An {\sc Anomaly} is signalled if the leader does not receive the requested bit value $b$ by the end of round~$t$.
\end{proof}

We conclude this subsection with a lemma showing that every RAM instruction can be efficiently emulated by an MC$+$ protocol.

\begin{lemma}\label{l:MC$+$_ram_instructions}
For every RAM instruction, there exists an MC$+$ protocol that whp correctly emulates it in at most $3w$ rounds; otherwise, an \textsc{Anomaly} is signalled.
The emulation requires $O(\rho'w)$ states per round for the leader and the agents in the working set.
\end{lemma}
\begin{proof}
The reductions in Subsection~\ref{ss:ram_instructions} and Fact~\ref{f:RAM_instructions}, together with Lemma~\ref{l:MC$+$_getput} and Theorem~\ref{th:combine_MC$+$}, imply that each RAM instruction can be emulated by an MC$+$ protocol in at most $3w$ rounds.
For an indirectly addressed register, the leader obtains its index using $\Call{SET}{}$ (Protocol~\ref{alg:set}).
Since this index can identify any of the $\rho'$ registers, its representation together with the current bit index requires $O(\rho'w)$ states per round for the leader.
The addresses of the remaining registers processed by the leader during this time appear explicitly in the program code and thus require only $O(1)$ states.
The agents in the working set also require $O(\rho'w)$ states per round to propagate requests.
\end{proof}

\subsection{\texorpdfstring{MC$+$}{MC+} Input and Output Conversion}\label{ss:MC$+$_inout}

As described in Subsection~\ref{ss:general_scheme}, the input values are converted from unary to binary and stored in the input array registers before the RAM instructions of the modified $\mathcal{P}$-program are emulated.
After the emulation, the values stored in the output array are converted back to unary to produce the protocol's output.
In this subsection we describe both conversions in the MC$+$ simulation.
We assume that a leader has been elected, the registers have been initialised, and the system of weights has been created.
Both conversions use this system of weights.

Let $X_i$ denote the subpopulation representing the $i$-th input variable, namely, the set of agents in state $in_i$.
Similarly, let $Y_i$ denote the subpopulation representing the $i$-th output variable, namely, the set of agents in state $out_i$.

We first describe input conversion (Protocol~\ref{alg:MC$+$_convert}).
The protocol builds on the mechanism used for parity computation in~\cite{gasieniec_et_al:LIPIcs.DISC.2026.17}, but recovers all bits of each input variable rather than only the least significant one, which determines parity.
The input variables are processed one after another.
To determine the binary representation of $|X_i|$, the protocol considers the weights in decreasing order, starting with $P_{w-2}$.
For each weight $P_\ell$, it adds that weight to an auxiliary subpopulation $Z$ and uses the MC$+$ majority protocol to determine whether $|Z|\le|X_i|$.
If the inequality holds, $P_\ell$ remains in $Z$ and the $\ell$-th bit of $|X_i|$ is set to $1$.
Otherwise, the weight is removed from $Z$ and the bit is set to $0$.
Each bit is stored in the input array register $r_i^{in}$ using $\Call{PUT\textsuperscript{e}}{}$.

Weights are added to and removed from $Z$ using $\Call{AddWeight}{}$ and $\Call{RemoveWeight}{}$, each of which completes whp within one round.
The protocol $\Call{AddWeight}{\ell}$ adds $P_\ell$ to $Z$ through a process initiated by the leader at the beginning of a new round.
Each agent entering this round receives information indicating that this process is in progress.
In particular, every agent in $P_\ell$ receives this information upon entering the round and joins $Z$.
If an agent skips this round, a clock anomaly is signalled.
The protocol $\Call{RemoveWeight}{\ell}$ is analogous, except that agents in $P_\ell$ leave $Z$, and $Z\gets\emptyset$ is shorthand for a similar subprotocol that removes all weights from $Z$ in one round.
Thus, in the absence of a clock anomaly, these operations are performed correctly.

\begin{algorithm}[H]
\begin{algorithmic}[1]
\Require{$i$ - input variable index}
\Ensure{Input array register $r_i^{in}$ stores $|X_i|$ in binary whp}
\Function{U2Bconversion\textsuperscript{e}}{$i$}
\State $Z\gets\emptyset$
\For{$\ell\gets w-2$ \textbf{downto} $0$}
  \State{\Call{AddWeight}{$\ell$} \Comment{Add weight $P_\ell$ to $Z$}}
  \State{Compare $X_i$ and $Z$ using $\Call{Majority}{}$.}
  \DLineIfElse{$|X_i|<|Z|$}{
  \State \hspace{\algorithmicindent} $\Call{RemoveWeight}{\ell}$ \Comment{Remove weight $P_\ell$ from $Z$}
  \State \hspace{\algorithmicindent} $\Call{PUT\textsuperscript{e}}{r_i^{in},\ell,0}$
  }
{\label{line:u2b_put1}
  \State \hspace{\algorithmicindent} $\Call{PUT\textsuperscript{e}}{r_i^{in},\ell,1}$
  }
\EndFor
\EndFunction
\end{algorithmic}
\caption{$\Call{U2Bconversion\textsuperscript{e}}{i}$}\label{alg:MC$+$_convert}
\end{algorithm}

The conversion of a single input variable is summarised by the following lemma.

\begin{lemma}\label{l:MC$+$_input_conversion}
The protocol $\Call{U2Bconversion\textsuperscript{e}}{i}$ correctly converts the $i$-th input variable from unary to binary and stores its binary representation in $r_i^{in}$ whp within $O(w\log n)$ rounds, using $O(1)$ additional states per round; otherwise, an \textsc{Anomaly} is signalled.
\end{lemma}

\begin{proof}
Assume that all component operations execute correctly. The greedy invariant in the binary-search discussion of Section~\ref{s:MC$+$_sim} proves that the value bits represent $A=|X_i|<2^{w-1}$ and the untouched sign bit is~$0$. Thus register $r_i^{in}$ stores the correct input value.

There are $w-1$ iterations, each using one majority comparison in
$O(\log n)$ rounds and a constant number of single-round weight operations
and bit writes. Including the initial clearing of~$Z$ gives
$O(w\log n)$ rounds. Theorem~\ref{t:MC$+$_majority},
Lemma~\ref{l:MC$+$_getput}, and the weight-operation guarantees provide
correctness whp and anomaly detection for each component.
Theorem~\ref{th:combine_MC$+$} gives the same guarantees for their
sequential composition. The components' additional state requirements
remain $O(1)$ per round.
\end{proof}

We now describe the conversion of the output array from binary to unary (Protocol~\ref{alg:MC$+$_unconvert}).
As with input conversion, the protocol processes the array sequentially, one entry at a time.

To convert the value stored in the $i$-th output register $r_i^{out}$, the protocol processes its bits one at a time.
For bit $\ell$, the leader first retrieves its value using $\Call{GET\textsuperscript{e}}{r_i^{out},\ell}$.
If the bit is $1$, the single-round protocol $\Call{Out}{\ell,i}$ places $2^\ell$ additional agents in state $out_i$.

\begin{algorithm}[H]
\begin{algorithmic}[1]
\Require{$i$ -- output variable index}
\Ensure{A set $Y_i$ of agents in state $out_i$ encodes the $i$-th output value in unary}
\Function{B2Uconversion\textsuperscript{e}}{$i$}
\For{$\ell\gets0$ \textbf{to} $w-2$}
  \State in round $2\ell$: $b\gets\Call{GET\textsuperscript{e}}{r_i^{out},\ell}$
  \DLineIf{$b=1$}{in round $2\ell+1$: $\Call{Out}{\ell,i}$}{}
\EndFor
\EndFunction
\end{algorithmic}
\caption{$\Call{B2Uconversion\textsuperscript{e}}{i}$}\label{alg:MC$+$_unconvert}
\end{algorithm}

The protocol $\Call{Out}{\ell,i}$ operates as follows (Protocol~\ref{alg:out}).
It uses weight $P_\ell$ of $2^\ell$ agents.
Each agent in $P_\ell$ adopts an additional component $out_i$ and transfers it to a free agent while retaining its weight label.
This adds $2^\ell$ agents to the output subpopulation.
After all bits of $r_i^{out}$ have been processed, the total number of agents in state $out_i$ equals the value stored in that register.

\begin{algorithm}[H]
\begin{algorithmic}[1]
\Require{$\ell$ -- weight index; $i$ -- output variable index}
\Ensure{$2^\ell$ additional agents enter state $out_i$}
\Function{Out}{$\ell,i$}
\State \textbf{upon} entering round $t$: \Comment{The protocol executes in one round}
\State \hspace{\algorithmicindent} the leader adopts component $(\ell,i)$;
\State \hspace{\algorithmicindent} a free agent adopts component $(\ell,i)$;
\State \hspace{\algorithmicindent} an agent in $P_\ell$ adopts component $out_i$, giving $(p_\ell,out_i)$.
\State \textbf{during} round $t$:
\State \hspace{\algorithmicindent} $(p_\ell,out_i)+\varphi\rightarrow p_\ell+out_i$. \Comment{$\varphi$ denotes a free agent}
\State \textbf{upon} entering round $t+1$:
\indLineIf{an agent still has $(p_\ell,out_i)$}{\textbf{signal} \textsc{Anomaly}}
\EndFunction
\end{algorithmic}
\caption{$\Call{Out}{\ell,i}$}\label{alg:out}
\end{algorithm}

\begin{lemma}\label{l:MC$+$_out}
The protocol $\Call{Out}{\ell, i}$ correctly creates a subpopulation of $2^\ell$ agents in state $out_i$ whp in one round, otherwise an \textsc{Anomaly} is signalled.
\end{lemma}
\begin{proof}
Let $t$ be the round in which $\Call{Out}{\ell,i}$ is executed.
Each agent in $P_\ell$ adopts an additional component $out_i$ upon entering round $t$.
For the protocol to succeed, each such agent must interact with a free agent during this round and transfer $out_i$ to that agent.
Throughout the protocol, at least $(1-\frac{1}{C})n$ agents remain free and form the working set.
Thus, for any agent in $P_\ell$ with $out_i$, the probability that a given interaction is initiated by this agent with a free agent as responder is at least $\frac{1-\frac{1}{C}}{n}$.
The phase clock ensures whp that all agents remain in round $t$ for at least $cn\log n$ interactions.
The probability that a particular agent retains $out_i$ while this common round lasts for at least $cn\log n$ interactions is at most
\[
\left(1-\frac{1-\frac{1}{C}}{n}\right)^{cn\log n}
\le e^{-(1-\frac{1}{C})c\log n}
\le n^{-(1-\frac{1}{C})c}.
\]
A union bound over the agents in $P_\ell$ gives a failure probability of at most
$
|P_\ell|\,n^{-(1-\frac{1}{C})c}\le n^{-\eta},
$
for any fixed $\eta>0$, provided that $c$ is sufficiently large.
Failure therefore requires either a common round shorter than $cn\log n$ interactions or an unsuccessful transfer within its first $cn\log n$ interactions.
Adding the phase clock's failure probability to this bound yields the whp guarantee.

\end{proof}

We conclude with a lemma on the binary-to-unary conversion of one output variable.

\begin{lemma}\label{l:MC$+$output_conv}
The protocol $\Call{B2Uconversion\textsuperscript{e}}{i}$ correctly converts the $i$-th output value stored in $r_i^{out}$ from binary to unary whp in $2w$ rounds, using $O(1)$ additional states per round; otherwise, an \textsc{Anomaly} is signalled.
\end{lemma}

\begin{proof}
Write the admissible output value as
$B=\sum_{\ell=0}^{w-2}b_\ell2^\ell<2^{w-1}$.
Initially, no agent belongs to~$Y_i$. If the component operations execute
correctly, then after processing bits $0,\ldots,\ell$, the set~$Y_i$
contains exactly $\sum_{j=0}^{\ell}b_j2^j$ agents: the bit retrieval
returns $b_\ell$, and $\Call{Out}{\ell,i}$ adds exactly $2^\ell$ agents
when this bit is~$1$, by Lemmas~\ref{l:MC$+$_getput} and~\ref{l:MC$+$_out}.
The population bound supplies the required free agents throughout.
Thus the final output count is~$B$, including $Y_i=\emptyset$ when $B=0$.

Each bit uses one retrieval round and at most one output round, giving
at most $2(w-1)<2w$ rounds. The same lemmas and
Theorem~\ref{th:combine_MC$+$} give correctness whp and anomaly detection
for the composition. Only the components' $O(1)$ additional states per
round are needed.
\end{proof}
\subsection{\texorpdfstring{MC$+$}{MC+} Simulation Summary}\label{ss:MC$+$summary}

We now combine the components developed in this section to prove Theorem~\ref{t:MC$+$sim}, which establishes the existence of an efficient MC$+$ simulation for any given $\mathcal{P}$-program.

\begin{proof}[Proof of Theorem~\ref{t:MC$+$sim}]
We follow the scheme in Protocol~\ref{alg:MC$+$_simulation_scheme}.
Each component is implemented by an MC$+$ protocol described above.
Since the number of component invocations is polylogarithmic, their composition is also an MC$+$ protocol by Theorem~\ref{th:combine_MC$+$}.

We first establish the round complexity.
Leader election takes $\Theta(w)$ rounds whp by Theorem~\ref{t:MC$+$_le}.
Register initialisation occupies one long round of $O(\rho'w)$ time whp by Lemma~\ref{l:reg_init}.
The system of $w-1$ weights is then created whp in one round by Lemma~\ref{l:weights_MC$+$}.
Next, each of the $T_R$ input values is converted from unary to binary and stored in the input array.
By Lemma~\ref{l:MC$+$_input_conversion}, each conversion takes $O(w\log n)=O(w^2)$ rounds, giving $O(T_Rw^2)$ rounds in total.
Emulating the modified $\mathcal{P}$-program takes $O(Tw)$ rounds by Lemma~\ref{l:MC$+$_ram_instructions}.
Finally, converting the $T_P$ output registers from binary to unary takes $O(T_Pw)$ rounds by Lemma~\ref{l:MC$+$output_conv}.
Terminal verification adds a constant number of rounds. Thus, in addition to the long register-initialisation round, the total number of ordinary rounds is
\[
\tau=O\!\left(w\,(T_Rw+T_{\mathrm{RAM}}+T_P)\right).
\]

We next establish the state complexity.
Leader election uses $O(w^2)$ states by Theorem~\ref{t:MC$+$_le}.
After register initialisation, the phase-clock component uses only $O(1)$ states per round by Theorem~\ref{t:clock}.
Weight creation uses $O(w^2)$ states by Lemma~\ref{l:weights_MC$+$}.
During subsequent computation, weight agents require $O(w\tau)$ states to record their weight labels and the current round.
Similarly, input agents require $O(T_R\tau)$ states and output agents require $O(T_P\tau)$ states.
Register agents require $O(\rho'w\tau)$ states to record their register indices, bit indices, bit values, and the current round.
The leader and the agents in the working set also require $O(\rho'w\tau)$ states.
For all agents, it suffices to use only $O(1)$ states per round for the remaining subprotocol-specific components.

Since $T_R,T_P\le\rho'$, the register-agent state bound dominates these contributions.
Thus, the total number of states is
\[
O(\rho'w\tau)
=O\!\left(\rho'w^2\,(T_Rw+T_{\mathrm{RAM}}+T_P)\right).
\]
\end{proof}

\noindent\textbf{Refined state complexity of the MC$+$ simulation.}
With additional technical work, which exploits the fixed order in which register bits are accessed, the state bound of Theorem~\ref{t:MC$+$sim}, and consequently that of Theorem~\ref{t:main}, can be improved by a factor of $\Theta(\min(\rho',w))$.

\tmcpr*

\begin{proof}
In the proof of Theorem~\ref{t:MC$+$sim}, the state complexity $O(\rho'w\tau)$ comes from the leader, the working set agents, and the register agents.
We first show, using three observations, that the leader and the working set need only $O(\rho')$ states per round.
We then show that the register agents, whose identities alone require $\Theta(\rho'w)$ states, need to count only $O(\tau/w)$ rounds.

\emph{Leader and working set agents.}
First, we fix the schedule of the simulation so that the bit index of every atomic subroutine call is determined by the round.
By Lemma~\ref{l:MC$+$_ram_instructions}, emulating a RAM instruction requires at most $3w$ calls to atomic subroutines, each in separate round.
If it requires fewer, we pad its emulation with idle rounds so that it occupies exactly $3w$ rounds.
In $\Call{U2Bconversion\textsuperscript{e}}{}$, the branch without $\Call{RemoveWeight}{}$ (line~\ref{line:u2b_put1} of Protocol~\ref{alg:MC$+$_convert}) is padded with one idle round, so that every iteration has the same number of rounds. The initial clearing of $Z$ is included in the first iteration's pass and, if necessary, padded to the same fixed pass length.
Since leader election has a fixed number of rounds, and register initialisation and weight creation take one round each, the input conversion starts and ends at fixed rounds.
Hence the indices of the input variable and of the bit are determined by the round number and the emulation of the RAM instructions also starts at a fixed round, depending only on $w$ and $T_R$.
We also pad $\Call{B2Uconversion\textsuperscript{e}}{}$, so that the round $2\ell+1$ is idle if bit~$\ell$ is~$0$, and the conversion of each output variable, which takes $2(w-1)<3w$ rounds, is padded to $3w$ rounds.
We call each of these fixed-length segments a \emph{pass}: an instruction slot or the conversion of one output variable, each of $3w$ rounds, or one iteration of an input conversion, of $\Theta(w)$ rounds. Execution is padded to all $T^*$ instruction slots before output conversion begins. Every pass has at least $w$ rounds and its boundaries are fixed in advance. The constant number of setup and terminal rounds use distinguished setup and terminal statuses.
The bit index of each $\Call{PUT\textsuperscript{e}}{rd,i,b}$ or $\Call{GET\textsuperscript{e}}{rs,i}$ call is therefore determined by the position of the round within its pass and by the current instruction or conversion step, so the leader and the working set agents need no separate component for it.
Second, register indices appearing explicitly in the instructions of the modified $\mathcal{P}$-program require only $O(1)$ states, since the program has fixed length.
Third, for an indirectly addressed register, the leader obtains its index using $\Call{SET}{}$ (Protocol~\ref{alg:set}).
Since this index may identify any of the $\rho'$ registers, its representation requires $O(\rho')$ states per round for the leader, and the agents in the working set also require $O(\rho')$ states per round to propagate requests involving this index.
Similarly, during output conversion, the leader stores the index of the current output variable, which requires $O(T_P)=O(\rho')$ states.
Hence the leader and the working set agents require only $O(\rho'\tau)$ states.

\emph{Register agents.}
Register agents require a different argument: their identities $(r,i,b)$ already account for $\Theta(\rho'w)$ states, which cannot be reduced, so we instead reduce the number of round values they store.
To this end, we exploit the fixed order in which the bits of each register are accessed.
Instead of counting every round, each register agent in state $(r,i,b)$ tracks \emph{coarse-grained rounds}, which advance once per pass.
It also stores the instruction emulated in the current pass, or a fixed tag for an idle or conversion pass, which requires only $O(1)$ states, since the program has fixed length.
The agent learns both from its interaction partners, whose states contain the current round and current instruction. The instruction tag remains fixed for the entire pass, including padding; an updated instruction pointer takes effect in the next pass.
Within a pass, the rounds in which bit~$i$ of any register may be accessed are determined by $i$ and the instruction, or, in a conversion, by $i$ and the coarse-grained round.
There are at most three of them, denoted $t_0<t_1<t_2$, and they may be consecutive (e.g.\ for $\Call{ADD}{r,r,r}$).
A register agent therefore also keeps a \emph{status} from the set
$\{early,\, active_0,\, between_0,\, active_1,\, between_1,\, active_2,\, late\}$.
It indicates whether the current round of the pass precedes $t_0$, equals $t_j$, lies strictly between $t_j$ and $t_{j+1}$, or follows the last of these rounds.
In addition, as in $\Call{PUT\textsuperscript{e}}{}$, the agent adopts the component $(put)$ when it enters an active round in which a $\Call{PUT\textsuperscript{e}}{}$ is executed, and clears it once it has processed the request.
Both components require $O(1)$ states.

\emph{Anomaly detection.}
These states suffice to preserve correctness and anomaly detection.
Register agents do not relay the fine clock or participate in majority operations. Both protocols run on the remaining $n-\rho'w=(1-o(1))n$ agents; all input and weight agents are in this set, so removing the register agents, which would contribute zero bias, does not alter a comparison. The induced scheduler is uniform on this set, and the constant-factor change in interaction rate preserves the bounds.
Outside its active rounds, no correct request concerns a register agent, so it ignores all requests.
Let $p$ denote the coarse-grained round of the agent, $I$ the instruction of the current pass, and $s$ its status; from $i$ and $I$, the agent knows the positions $t_0<t_1<t_2$ of its active rounds within the pass.
When it meets a non-register agent in fine round $t'$ with instruction or pass tag $I'$, the transition function determines the corresponding coarse-grained round $p'$ and status $s'$ from $t'$, $I'$, and~$i$. This is possible because the pass boundaries are fixed. Coarse-grained rounds and statuses within a pass advance monotonically; an older observation is ignored, except for the anomaly tests below. Two different instruction tags for the same pass signal an anomaly. Before replacing $(p,I,s)$ by a later $(p',I',s')$, the agent checks all obligations associated with the old status and every active round that the update would skip. Thus an update never erases an unchecked witness. The exact round $t'$ is not stored.
The agent signals an anomaly in the following cases:
\begin{enumerate}
    \item \emph{Status skip:} the update would skip one of its active rounds, that is, either $p'=p$ and the next active round of the agent in pass $p$ precedes $t'$, or $p'=p+1$ and the agent has an active round in pass $p$ after its current status or in pass $p'$ before $t'$.
    \item \emph{Clock anomaly:} the partner's coarse-grained round differs from $p$ by more than one, or the agent is active, the partner carries a fine round number, and that number differs by more than one from the active round determined by $p$, $I$, $i$, and~$s$.
    \item \emph{Protocol anomaly:} the agent leaves an active status of a $\Call{PUT\textsuperscript{e}}{}$ round while still holding the component $(put)$.
\end{enumerate}
In active rounds in which $\Call{GET\textsuperscript{e}}{}$ is executed, the agent adopts the component $(get)$ as in Protocol~\ref{alg:get}, which allows it to answer the request; no additional check is attached to it, since the leader signals an anomaly if it does not receive the bit. Requests and responses are accepted only for the corresponding active round. On receipt of the terminal marker, the agent checks every remaining active-round obligation and any pending $(put)$ component before entering its terminal status. An anomaly indication is permanent and takes priority over subsequent updates.
Since every round contains an interval of $\Omega(\log n)$ time in which all other agents share the same round number (Theorem~\ref{t:clock}), a register agent enters each of its active statuses whp, so none of these anomalies occurs in a correct execution whp.

Since every pass has at least $w$ rounds, there are $O(\tau/w)$ coarse-grained rounds.
A register agent thus stores its register index, bit index, and bit value, its coarse-grained round, and $O(1)$ further states for the instruction, the status, and the components $(put)$ and $(get)$.
Hence the register agents use $O(\rho'w\cdot\tau/w)=O(\rho'\tau)$ states instead of $O(\rho'w\tau)$.
The remaining bounds from the proof of Theorem~\ref{t:MC$+$sim} are unchanged: leader election and weight creation use $O(w^2)$ states, weight agents use $O(w\tau)$ states, and input and output agents use $O((T_R+T_P)\tau)=O(\rho'\tau)$ states. Hence the total number of states is $O((\rho'+w)\tau)$. The backup contributes $O(\rho'w^2)$ states, which is absorbed because $\rho'=O(T)$ and $\tau=\Omega(wT)$ for the padded schedule. Thus the same refinement applies to the combined stable and silent simulation.
\end{proof}

\section{Backup Simulation}\label{s:stable_sim}\label{app:backup}

\subsection{Overview}\label{ss:backup_overview}

We now present a novel stable and silent backup protocol simulating general $\mathcal{P}$-programs, which provides the stable component required by Theorem~\ref{th:stable+MC$+$}.
An anomaly propagates a one-way switch into fresh backup states, preserving input labels. The backup then follows Figure~\ref{fig:sim_scheme} with the same instruction reductions (Subsection~\ref{ss:ram_instructions}), without a clock or further epidemics; full details and proofs follow in this section.

\begin{restatable}{theorem}{tbackup}\label{t:stable_sim}
For every $\mathcal{P}$-program, the backup simulation population protocol silently stabilises in expected time $O(n^2w)$, and uses $O(\rho'w\log n)=O(\rho'w^2)$ states.
\end{restatable}

A stable protocol must cope with two problems that the MC$+$ simulation sidesteps by failing with negligible probability: the eventual leader never learns that it is unique, and it never learns that it has read all input agents.
The construction addresses them with three main techniques.

\noindent\textbf{(1) Rank-based restarts.}
Leaders are eliminated in pairwise duels~\cite{AAD+06}, which we augment with a \emph{rank} carried by every agent; in a state $(X,i)$, $X\in\{L,F\}$ marks a leader or a non-input follower and $i$ is its rank:
\[
(L,i)+(L|F,j)\rightarrow (L,i)+(F,i)\ \text{ if } i>j,\qquad (L,i)+(L,i)\rightarrow (L,i+1)+(F,i+1).
\]
This augmentation is essential. Without it, simulations started by leaders that are later eliminated leave register, output and read-input data that the final leader cannot distinguish from its own, and since it never learns that it is unique, it cannot know when to start the definitive simulation.
The rank of a leader is the Strahler number~\cite{Stra52} of its tree of won duels, so it is at most $1+\lfloor\log n\rfloor$, and only the finally elected leader reaches the highest rank (Lemma~\ref{l:le_and_rank}).
After each rank increase, the leader restarts the entire simulation under its new rank. Rank rules take priority throughout; a lower-ranked non-input follower can be reclaimed and, when needed, allocated in the same interaction.
Hence the simulation of the last leader overwrites all others and alone determines the output, although this leader never learns that it is unique.
Unlike a counter of won duels, which could reach~$n$, the rank stays logarithmic, so ranks replace the round counter of the MC$+$ simulation at a cost of only a factor $O(\log n)=O(w)$ in the number of states.

\noindent\textbf{(2) Direct access.}
In $\Call{PUT\textsuperscript{b}}{}$ and $\Call{GET\textsuperscript{b}}{}$ (stable variants of the atomic subroutines), the leader simply waits until it meets the agent holding the requested register bit, which happens with probability~$1$ in $O(n)$ expected time.
Once the unique highest-ranked leader remains, the reductions of Subsection~\ref{ss:ram_instructions} emulate every RAM instruction correctly with probability~$1$ in $O(nw)$ expected time (Fact~\ref{f:stable_getput}); program emulation and output conversion then always succeed (Lemma~\ref{l:stable_B2U}).

\noindent\textbf{(3) Self-correcting input reading.}
Input agents keep their states $in_i$ permanently; the leader increments $r_i^{in}$ for every input agent of lower rank that it meets and marks it as read by assigning it its own rank.
Termination of the input conversion stage is decided by a \emph{helper} agent that counts its interactions with the leader up to $dw$: as the leader meets a given unread input agent and the helper with equal probability, it misses some input agent with probability at most $n2^{-dw}\le n^{-\eta}$ (Lemma~\ref{l:stable_U2B}).
This is where the preprocessing of Subsection~\ref{ss:general_scheme} is essential.
Since the input array is protected and the output array is preserved during printing, a missed input agent can be handled after program emulation and output conversion finish: the leader erases the completed unary output (Lemma~\ref{l:stable_B2U}) and re-enters input conversion. Only previously missed input agents remain unread, so their counts are added to the retained input array.
The program is then emulated again on the corrected input array.

After leader election, each fresh input-conversion attempt misses an input agent with conditional probability at most $n^{-\eta}$, so the expected number of further retries is $o(1)$. Any attempt already in progress has the same asymptotic expected cost.
The emulation takes $O(Tnw)=O(n\poly(w))$ expected time, so the bound $O(n^2w)$ is dominated by the conversions of the $O((T_R+T_P)2^{w-1})=O(n)$ unary input and output units, each requiring an $\Call{INC}{}$ or $\Call{DEC}{}$ of $O(nw)$ expected time.

\subsection{State Organisation and Handover}\label{ss:backup_handover}
The following subsections give the full description and proofs of the backup simulation outlined in Subsection~\ref{ss:backup_overview}.
\bigskip

In this section, we present a stable and silent backup protocol for simulating a given $\mathcal{P}$-program (Protocol~\ref{alg:stable_sim}).
The program halts on every admissible input and initialises each working register before its first read in that execution, including on the partial input vectors encountered during recovery. Its time, memory, and input/output bounds hold uniformly for these executions.
It provides the stable component required by the MC$+$ framework.
The construction silently stabilises to the correct output with probability~$1$, in expected time $O(n^2w)$, using $O(\rho'w\log n)=O(\rho'w^2)$ states.
As in the MC$+$ simulation, a leader coordinates the computation, as described in Subsection~\ref{ss:general_scheme}.
However, in the backup, the leader accesses the register, input, and output agents directly.
Other agents store data or assist with input conversion; they do not propagate requests by epidemics. Rules involving the leader apply in both interaction orientations.

When an MC$+$ anomaly invokes this backup, a one-way backup-mode flag spreads by epidemic. On first receiving the flag, a non-input agent discards its MC$+$ state and enters the backup as a rank-$1$ leader, while an input agent retains its immutable input label and receives rank~$0$. An agent already in backup mode is never initialised again. The rank rules below remain active during this handover. All agents enter backup mode in expected time $O(\log n)$; the ensuing analysis includes any simulation work already in progress. Thus the handover needs no simultaneous reset and uses disjoint MC$+$ and backup state spaces.

The backup does not count rounds.
Instead, each agent carries a \emph{rank}, which distinguishes successive simulations and allows obsolete data to be discarded.
In protocol descriptions, agent $A$ usually denotes a leader and agent $B$ its interaction partner.
The state of an agent $A$ is a tuple of components, one of which is always the rank $A.rank$.
We refer to the remaining components collectively as the \emph{substate} $A.substate$.
When a subprotocol has an explicit argument $A$, this denotes the leader executing it.
The component $rank$ takes values in $\{0,1,\ldots,h\}$, where $h=1+\lfloor\log n\rfloor=O(w)$, while $substate$ specifies the role of the agent and its role-specific data, as listed in Table~\ref{tab:fields}.

\begin{table}[H]
\centering
\begin{tabular}{l||l||l}
\textbf{Role} & \textbf{Value of $substate$} & \textbf{Description} \\\hline\hline
leader & $(L,aux)$ & $aux$: finite control, $ip$, one register index, \\
       &           & one bit index, input/output index\\
register & $(R,r,i,b)$ & register index, bit index, bit value \\
helper & $(H,j)$ & counter of interactions with the leader \\
input & $in_i$ & index of the input variable \\
output & $out_i$ & index of the output variable \\
free & $\varphi$ & no assigned role \\
\end{tabular}
\caption{Values of the component $substate$.}
\label{tab:fields}
\end{table}

\subsection{Ranks, Leader Election and Backup Simulation Overview}

The first stage of the backup simulation is leader election and rank computation. Election and rank-reclamation rules take priority in every control state, including during subroutine calls and after output conversion.
We use a standard leader election protocol based on \emph{duels}, augmented with a \emph{rank-computation mechanism} that stores the rank of each agent in its component $rank$.

In this stage, the transition function concerns only the role and rank component.
On initial entry into the backup, each non-input agent is a leader~$L$ of rank~$1$, and each input agent has rank~$0$.
Whenever two leaders interact, one of them is eliminated: if their ranks differ, the leader with the lower rank becomes a follower $F$; if the ranks are equal, one of them becomes a follower and the other increases its rank by~$1$.
Only non-input agents reach follower state $F$.
Consequently, only one leader reaches the highest rank that is at most $1+\lfloor\log n\rfloor$.
A leader reclaims a lower-ranked non-input partner by resetting it to the free substate at its own rank. When the current operation allocates a free agent, a lower-ranked non-input follower may be reclaimed and assigned its new role in the same interaction; leader duels retain priority. An input agent retains its immutable input label and receives the leader's rank only when accepted for input conversion. Every other simulation operation on a non-input agent requires matching ranks; higher-ranked partners are left unchanged.
This process is described by the following transitions:
\[(L,i)+(L|F,j)\rightarrow (L,i)+(F,i)\;\mbox{ if }\; i>j,\]
\[(L,i)+(L,i)\rightarrow (L,i+1)+(F,i+1).\]
Here $F$ denotes a non-input follower; on losing a duel, it is reset to the free substate. In an equal-rank duel, the winner discards its old control state and restarts register allocation at the increased rank. Interactions between two non-leaders make no changes. Once a unique leader of the highest rank remains, which happens with probability~$1$, it reclaims the remaining lower-ranked non-input agents through direct interactions.

\begin{lemma}\label{l:le_and_rank}
The leader election and rank computation stage of the backup simulation elects a unique leader in expected time $O(n)$.
Only this leader reaches the highest rank in the population, which is at most $1+\lfloor\log n\rfloor$.
\end{lemma}
\begin{proof}
The time complexity of leader election by pairwise duels has been studied
in~\cite{AAD+06}.
The rank of a leader corresponds to the {\em Strahler number}~\cite{Stra52}
of the tree of duels it has won, or equivalently to the {\em register
number}~\cite{FlajoletRV79,EsparzaLS14}. In particular, a leader of rank
$r$ has defeated, directly or indirectly, at least $2^{r-1}$ initial leaders,
so $2^{r-1}\le n$.
No other agent ever holds the leader role with the final highest rank: if two agents ever reached that rank as leaders, eliminating either would require an equal-rank duel creating a higher rank or a duel with an already higher-ranked leader. The maximum leader rank never decreases, contradicting its final value. These arguments also allow staggered entry during handover: each non-input agent contributes once as an initial rank-$1$ leader, and the $O(n)$ expected election bound applies after the last such entry.
\end{proof}

After each rank increase, the leader restarts the entire simulation and labels it with its new rank, thereby invalidating all simulations of lower ranks.
Hence, the simulation started by the last remaining leader is the only one with the highest rank; it overwrites all previous simulations, and its result becomes the final output of the backup protocol.
This mechanism allows the finally elected leader to complete the definitive simulation without ever receiving confirmation that it is the only leader. When it first attains its final rank, that rank has never labelled another leader's simulation. The register representation it constructs is therefore fresh and cannot be modified by lower-ranked leaders.

The input array is cumulative and protected: the source program addresses only its original $\rho$ registers, and only the trusted replacements of $\Call{READ}{}$ and $\Call{PRINT}{}$ access the reserved input and output arrays. Scratch registers are initialised before use. On a retry at the same rank, register agents and input counts are retained; the modified program restores its input/output pointers and initialises its working values. Address bounds are checked before an index is retained by the leader. An invalid request suspends that simulation while leaving the priority rank rules active; it cannot occur in the final isolated instance under the program assumptions.

\begin{algorithm}[H]
\begin{algorithmic}[1]
\Require{~a modified  $\mathcal{P}$-program, a leader $A$}
\Ensure{~stable simulation if $A$ is the unique leader}
\Function{Simulation\textsuperscript{b}}{$A$}

\State $\Call{RegInit}{\rho', w, A}$
\State  \textbf{U2B:} $\Call{U2Bconversion\textsuperscript{b}}{A}$
\State $ip\gets 0$
\While{$ip\neq \mathsf{END}$}
\State $A$ emulates the RAM instruction indicated by $ip$.
\State Update $ip$.
\EndWhile
\State $\Call{B2Uconversion\textsuperscript{b}}{A}$
\State \textbf{FINISHED:} retain the output and wait for an input agent $B$ with $B.rank<A.rank$.
\State $\Call{EraseOutput}{A}$; \textbf{goto U2B:}

\EndFunction
\end{algorithmic}
\caption{$\Call{Simulation\textsuperscript{b}}{}$. The final checking state persists; priority rank rules remain active throughout.}\label{alg:stable_sim}
\end{algorithm}

During its simulation, the leader uses only agents of its own rank; whenever it encounters followers of lower rank, it turns them into free agents.
Thus, unless stated otherwise, all agents appearing in the protocols executed by a leader~$A$ are assumed to have rank $A.rank$.
The leader starts the simulation by initialising the registers.
It then reads the input using the unary-to-binary conversion, which succeeds whp.
It then emulates the modified $\mathcal{P}$-program and converts the output array from binary to unary, both of which always succeed.
However, with negligible probability, some input agents are not read during the unary-to-binary conversion.
To handle this case, after completing the simulation, the leader keeps checking for input agents of lower rank than its own.
If there are none, these checks do not cause meaningful interactions and thus do not prevent the protocol from being silent.
If such an agent is found, the entire output is erased and the simulation returns to the unary-to-binary conversion stage to read the remaining input agents.
The working registers need not be reset externally before emulation is repeated, since the $\mathcal{P}$-program initialises them before reading them. The input array is preserved and subsequent conversions count only input agents of lower rank.
Since the modified $\mathcal{P}$-program performs the same number $T_P$ of prints for any input, there is also no need to reset the output array: it is overwritten when the emulation is restarted.

\subsection{Backup Emulation of RAM Instructions}\label{ss:stable_putget}

We begin with the stable backup variants $\Call{PUT\textsuperscript{b}}{}$ and $\Call{GET\textsuperscript{b}}{}$ of the atomic subroutines $\Call{PUT}{}$ and $\Call{GET}{}$.
In the $\Call{PUT\textsuperscript{b}}{rd, i, b}$ subroutine, the leader interacts with agents one after another until it encounters the agent holding the $i$-th bit of register $rd$, at which point it sets this bit to $b$.

\begin{algorithm}[H]
\begin{algorithmic}[1]
\Require{$rd$ - pointer to a register, $i$ - bit index, $b\in \{0, 1\}$, a leader $A$}
\Ensure{The $i$-th bit of register $rd$ is set to $b$}
\Function{PUT\textsuperscript{b}}{$rd, i, b, A$}
    \State \textbf{repeat}
    \State The leader interacts with an agent $B$.
    \State \textbf{until} $B.rank=A.rank$ \textbf{and} $B.substate=(R,rd,i,\star)$
    \State $B.substate \gets (R,rd,i,b)$
\EndFunction
\end{algorithmic}
\caption{$\Call{PUT\textsuperscript{b}}{rd, i, b}$}\label{alg:sput}
\end{algorithm}

The $\Call{GET\textsuperscript{b}}{rs, i}$ subroutine is analogous. The leader interacts with agents until it encounters the agent holding the $i$-th bit of register $rs$, at which point it learns the value of that bit.

\begin{algorithm}[H]
\begin{algorithmic}[1]
\Require{$rs$ - pointer to a register, $i$ - bit index, a leader $A$}
\Ensure{The leader learns the $i$-th bit of register $rs$}
\Function{GET\textsuperscript{b}}{$rs, i, A$}
    \State \textbf{repeat}
    \State The leader interacts with an agent $B$.
     \State \textbf{until} $B.rank=A.rank$ \textbf{and} $B.substate=(R,rs,i,b)$
    \State \textbf{return} $b$
\EndFunction
\end{algorithmic}
\caption{$\Call{GET\textsuperscript{b}}{rs, i}$}\label{alg:sget}
\end{algorithm}

Each of these subroutines completes once the leader interacts with the agent holding the specified bit.
Such an interaction occurs with probability~1, in expected time $O(n)$.
The following fact states the completion time of the backup variants of the atomic subroutines and, consequently, the time required to emulate RAM instructions.

\begin{fact}\label{f:stable_getput}
    When executed by the unique highest-ranked leader, the backup subroutines $\Call{GET\textsuperscript{b}}{}$ and $\Call{PUT\textsuperscript{b}}{}$ complete in expected time $O(n)$.
    Consequently, every RAM instruction can be stably emulated by the unique leader in expected time $O(nw)$.
\end{fact}

\subsection{Backup Input and Output Conversion}

In this subsection, we describe input and output conversion in the stable backup variant.
Since the input and output arrays occupy fixed ranges of consecutive addresses, with $r_i^{in}$ and $r_i^{out}$ at fixed positions,
we assume that the leader can convert an index $i$, or a state $in_i$ or $out_i$, to the address of $r_i^{in}$ or $r_i^{out}$, and vice versa.
This mapping can be encoded in the transition function, or emulated by a sequence of RAM instructions using a few additional registers.
\smallskip

\noindent\textbf{Backup unary-to-binary input conversion.}
Before the RAM instructions of the modified $\mathcal{P}$-program are emulated, the input data is converted from unary to binary and stored in the input array, as in the MC$+$ variant.
We now introduce the protocol $\Call{U2Bconversion\textsuperscript{b}}{A}$ (Protocol~\ref{alg:stable_convert}), executed by the leader~$A$, which performs this conversion.

First, the leader designates a free agent $B$ as a helper by setting $B.substate\gets(H,0)$.
The helper increments its counter each time it interacts with the leader.
The conversion protocol terminates when the helper's counter reaches the value $dw$, for a constant $d>0$.
At that point, the leader turns the helper back into a free agent so that every call of the protocol uses a fresh helper.

After passing the label \textbf{START:}, the leader is in a \emph{waiting} state, in which it reacts only to new input agents and the helper.
When it encounters an input agent $B$ with $B.substate=in_i$ and a lower rank, the leader executes line~\ref{line:new_input}: it marks $B$ as read by assigning it its own rank and increments $r_i^{in}$ by one.
Until this incrementation is completed, the leader ignores the helper and other input agents; only the required register accesses and the priority rank rules remain active.
Once the incrementation is completed, the leader returns to the waiting state and continues to process input agents and count interactions with the helper.

The helper's counter bounds the number of $L+H$ interactions and thus determines when the conversion ends.
By choosing a sufficiently large constant $d$, the leader $A$ encounters every input agent whp before the helper's counter reaches $dw$.

\begin{algorithm}[H]
\begin{algorithmic}[1]
\Require{~$A$ - a leader}
\Ensure{~the cumulative input counts include every input agent whp}
\Function{U2Bconversion\textsuperscript{b}}{$A$}
\State In the first interaction of $A$ with a free agent $B$:
\State \hspace{\algorithmicindent} $B.substate \gets (H,0)$ \Comment{Helper designation.}
\State \textbf{START:}
\State During subsequent interactions $A+B$:
\indLineIf{$B.substate=in_i$ \textbf{and} $B.rank<A.rank$}
{
\State \hspace{\algorithmicindent}\hspace{\algorithmicindent} $B.rank \gets A.rank, \; \Call{INC}{r^{in}_{i}}, \;$ \textbf{goto START:}\label{line:new_input}
}
\indLineIf{$B.substate=(H,j)$ \textbf{and} $B.rank=A.rank$}
{
  \indindLineIfElse{$j+1<dw$}{
  $B.substate \gets (H,j + 1), \;$ \textbf{goto START:}
  }{
  $B.substate\gets \varphi$; \textbf{return}
  }
}
\EndFunction
\end{algorithmic}
\caption{$\Call{U2Bconversion\textsuperscript{b}}{A}$}\label{alg:stable_convert}
\end{algorithm}

The following lemma characterises the unary-to-binary input conversion protocol of the backup.

\begin{lemma}\label{l:stable_U2B}
Protocol $\Call{U2Bconversion\textsuperscript{b}}{A}$, when executed by the unique leader of the final rank, returns in expected time $O((N+1)nw)=O(n^2w)$, where $N$ is the number of input agents not yet counted at the start of the call.
The probability that any input agent remains uncounted at return is at most $N2^{-dw}\le n2^{-dw}$.
\end{lemma}

\begin{proof}
After leader election, a constant fraction of the population is either free at the leader's rank or can be reclaimed from a lower rank on contact. Thus the leader designates a helper in expected time $O(1)$.

Fix an input agent $B$ not yet counted and consider only interactions of the leader in the waiting state. Until $B$ is counted, the probabilities of meeting $B$ and the helper are equal. Pausing the scan during register increments does not change this symmetry. Hence the probability of meeting the helper $dw$ times before meeting $B$ is $2^{-dw}$. A union bound over the $N$ uncounted input agents gives $N2^{-dw}$. This is at most $n^{-\eta}$ for any prescribed constant $\eta>0$ when $d$ is sufficiently large, since $w=\Theta(\log n)$.

The expected total time spent waiting for the $dw$ helper meetings is $O(nw)$. At most $N$ input-register increments are performed, each in expected time $O(nw)$ by Fact~\ref{f:stable_getput}. The helper counter is paused during these increments, giving total expected time $O((N+1)nw)$.
At every boundary between completed increments, each input register equals the number of input agents of its label that have been assigned the leader's rank. Thus partial conversion preserves correct cumulative counts.
\end{proof}

\noindent\textbf{Backup binary-to-unary output conversion.}
The results of the modified $\mathcal{P}$-program are stored in binary form in the output array.
We now describe the binary-to-unary conversion protocol in the stable backup variant $\Call{B2Uconversion\textsuperscript{b}}{A}$ (see Protocol~\ref{alg:stable_unconvert}), executed by the leader~$A$.

The leader executes a loop from $i=1$ to $T_P$.
Its goal in iteration $i$ is to create a subpopulation of $[r_i^{out}]$ agents in substate $out_i$, which represents the $i$-th output value in unary.
To this end, the leader copies $[r_i^{out}]$ into a temporary register $rc$ using the direct copy $\Call{COPY}{r_i^{out},rc}$.
It repeatedly decrements $rc$ and assigns the substate $out_i$ to a free agent whenever the resulting value is non-negative, stopping when $[rc]=-1$. Thus it prints exactly one agent per unit of the original output value, including no agents when that value is zero.

In this way, the computation results are output in unary form, while the original binary values remain stored in the output array.
Preserving these values allows the output to be undone if the simulation was carried out on incorrect data.
This can occur only if the leader fails to encounter some input agent(s) during the unary-to-binary input conversion stage.

After output conversion has completed, $\Call{EraseOutput}{A}$ follows the same scheme using the preserved binary values. Only after erasure finishes may the next execution overwrite the output array.
The only difference is that the leader, instead of turning agents in substate $\varphi$ into agents in substate $out_i$, turns agents in substate $out_i$ into agents in substate $\varphi$.

\begin{algorithm}[H]
\begin{algorithmic}[1]
\Require{~$A$ - a leader}
\Ensure{~all output stored in binary in output array registers stably converted to unary}
\Function{B2Uconversion\textsuperscript{b}}{$A$}
\For{$i\gets1$ \textbf{to} $T_P$}
\State $\Call{COPY}{r_i^{out}, rc}$
\State \textbf{goto CONVERT:}
\State \textbf{ASSIGN:}
\State In the first interaction of $A$ with a free agent $B$:\label{line:give_out1}
\State \hspace{\algorithmicindent} $B.substate \gets out_i$\label{line:give_out2}
\State \textbf{CONVERT:}
\State $\Call{DEC}{rc}, \; \Call{JGE}{rc, \textbf{ASSIGN:}}$
\EndFor
\EndFunction
\end{algorithmic}
\caption{$\Call{B2Uconversion\textsuperscript{b}}{A}$}\label{alg:stable_unconvert}
\end{algorithm}

The behaviour of the output-creating and output-erasing protocols is characterised by the following lemma.

\begin{lemma}\label{l:stable_B2U}
    Protocol $\Call{B2Uconversion\textsuperscript{b}}{A}$, when executed by the unique highest-ranked leader, stably converts all output values stored in the output array from binary to unary representation in expected time $O(T_P2^wnw)=O(n^2w)$.
    Moreover, if necessary, the unary output representation can be stably erased by protocol $\Call{EraseOutput}{A}$ in the same expected time.
\end{lemma}

\begin{proof}
By Fact~\ref{f:stable_getput}, each RAM instruction used in $\Call{B2Uconversion\textsuperscript{b}}{A}$ executes in expected time $O(nw)$.
After leader election, a constant fraction of the population is free or reclaimable on contact, even while the definitive output is being allocated.
Thus, the leader $A$ can stably assign any output state to a previously free agent in expected time $O(1)$ (lines~\ref{line:give_out1}--\ref{line:give_out2} of Protocol~\ref{alg:stable_unconvert}).
The binary-to-unary conversion proceeds one output variable at a time.
To convert the $i$-th output value, the leader emulates one instruction $\Call{COPY}{}$, followed by at most $2^{w-1}$ pairs of $\Call{DEC}{}$ and $\Call{JGE}{}$ instructions and exactly $[r_i^{out}]$ assignments of an output state to a free agent. Since $0\le[r_i^{out}]<2^{w-1}$, the signed counter runs down to $-1$ without overflow.
Each such iteration takes expected time $O(nw)$.
Hence, converting a single output variable takes expected time $O(2^w\cdot nw)$.
Consequently, all $T_P$ output variables are stably converted from binary to unary in total expected time $O(T_P2^wnw)=O(n^2w)$.

The analysis of protocol $\Call{EraseOutput}{A}$ is analogous to the one above.
The only difference is that releasing an output agent takes expected time up to $O(n)$.
However, this does not affect the overall expected time bound.
\end{proof}

\subsection{Backup Summary}

We conclude the construction of the backup simulation protocol by establishing the following theorem on its correctness and complexity.

\tbackup*

\begin{proof}
After the $O(\log n)$ expected handover, Lemma~\ref{l:le_and_rank} gives a unique leader within further expected time $O(n)$. Only this leader ever attains its final rank~$K$, so its rank-$K$ registers are uncontaminated. By the coupon collector bound, direct interactions reclaim all lower-ranked non-input agents in expected time $O(n\log n)$. Reclamation overlaps simulation; no separate synchronisation is needed.

At rank $K$, the protected input array records correct cumulative counts for agents assigned that rank and survives retries. Every tentative input is admissible: the program halts, initialises its working registers, and writes exactly $T_P$ outputs. Output conversion completes with probability~$1$, preserving the binary record needed for erasure. If input was missed, the leader eventually encounters a missed agent in \textbf{FINISHED}, erases the completed output, and resumes conversion with a fresh helper. Otherwise, the output is correct and no retry occurs.

Register allocation costs expected time $O(\rho'w)$ once at the final rank: a constant fraction of agents is free or reclaimable on contact, and performing reclamation and allocation in one interaction allows Lemma~\ref{l:reg_init} to apply. Lemma~\ref{l:stable_U2B}, Fact~\ref{f:stable_getput}, and Lemma~\ref{l:stable_B2U} give expected costs $O(n^2w)$ for input conversion, $O(Tnw)$ for each tentative execution, and $O(n^2w)$ each for output creation and erasure. Finding a missed input in \textbf{FINISHED} costs expected time $O(n)$. Since $T=\poly(w)$ and $n\ge C2^w$, these costs, including detection and erasure after failure, are bounded conditionally on every valid attempt-start history by
\[
B=O(n^2w).
\]
The same bound covers a legitimate subprotocol already in progress when leader election finishes. Charge $O(B)$ for this residual execution, then consider attempts starting with a fresh helper.

For each fresh attempt, Lemma~\ref{l:stable_U2B} bounds the conditional probability of missed input by
\[
\delta\le n2^{-dw}\le n^{-\eta},
\]
for sufficiently large $d$. Thus a $k$-th fresh attempt is needed with probability at most $\delta^{k-1}$. Summing the conditional cost bounds at attempt starts gives expected remaining time at most
\[
\sum_{k\ge1}\delta^{k-1}B=\frac{B}{1-\delta}=O(n^2w).
\]
Independence is unnecessary, and infinitely many retries have probability zero. After complete input has been simulated, obsolete non-input states are reclaimed, the helper is released, and the leader's checking rules make no changes. The protocol therefore silently stabilises correctly with probability~$1$. Including election, reclamation, allocation, and the initial residual execution preserves the expected bound $O(n^2w)$.

There are $O(\log n)=O(w)$ ranks. Per rank, register agents use $O(\rho'w)$ states for a register index, bit index, and bit value; input/output agents use $O(T_R+T_P)$; the helper uses $O(w)$; and free agents use $O(1)$. The leader stores $ip$ and finite control in $O(1)$ states for the fixed program, plus one variable register index, one bit index, and constantly many bits. During conversion, the input/output index determines the register address. Whole-word temporaries occupy auxiliary registers already included in $\rho'$. Thus the leader uses $O(\rho'w)$ states per rank, and the disjoint role-specific state spaces total $O(\rho'w\log n)$ states.
\end{proof}

\noindent\textbf{Remark.}
Correctness of the backup does not require advance knowledge of $T_{\mathrm{RAM}}$. It is enough that each admissible execution halts and initialises its working registers, while the memory and input/output bounds are known so that the required registers and agents can be allocated. The stated complexity bounds continue to use the polynomial resource assumptions defining $\mathcal{P}$-programs.

\section{Conclusions}\label{s:conclusions}

We have shown that every $\mathcal{P}$-program satisfying the stated halting and register-initialisation assumptions admits an efficient, stable, and silent population protocol (Theorem~\ref{t:main}). Under suitable binary encodings and population-capacity bounds, this supports computations in $\mathsf{P}$ and $\mathsf{FP}$, including multiple outputs; longer results can be split into admissible words. The total binary input length is $O(T_Rw)$ and need not equal~$w$. In particular, multiplication of input counts admits an efficient stable protocol with unary inputs and outputs whenever the product lies in the admissible range. This gives a lower bound towards the time-bounded characterisation sought in~\cite{CzernerFischerGuttenberg25}.

\emph{Las Vegas computations.} The framework suggests an extension to $\mathsf{ZPP}$. An instruction $\Call{LoadRandom}{rd}$ could sample a word bit by bit from the leader's initiator/responder roles at fresh interactions. If a zero-error algorithm has expected running time at most $p(w)$, each fresh trial truncated after $2p(w)$ steps succeeds with probability at least~$1/2$; $cw$ trials fail with probability at most $2^{-cw}=n^{-\Omega(1)}$. Formalising random-bit generation, restarts, and a stable backup remains necessary. Exhausting all trials would trigger fallback; this is distinct from an MC$+$ simulation anomaly.

\emph{Succinct Presburger evaluation.}
Czerner et al.~\cite{DBLP:journals/jcss/CzernerGHE24} proposed as a research direction succinct protocols for quantifier-free Presburger formulas, with $\poly(m)$ states for a formula of size~$m$, that converge in $\poly\log n$ time.
We propose an analogous direction for leaderless protocols that are additionally required to be stable.
With $\poly(m)$ states independent of~$n$, such protocols cannot be fast~\cite{BDS17}, and already majority, a formula of constant size, requires $\Omega(\log n)$ states for time $O(n^{1-c})$ under the assumptions of~\cite{DBLP:conf/soda/AlistarhAG18}.
We therefore call a fast leaderless stable protocol succinct if it uses $\poly(m)\cdot\poly\log n$ states.
Under this notion, our result gives succinct protocols that moreover stabilise silently: every such formula is evaluated by a $\mathcal{P}$-program with $\poly(m)$ instructions and registers, so for all sufficiently large~$n$, Theorem~\ref{t:main} yields a protocol with $\poly(m)\cdot\poly\log n$ states and $\poly\log n$ stabilisation time.
Reducing the polylogarithmic factor, ideally to $\poly(m)\cdot O(\log n)$ states, which would match the lower bound for majority, remains open.

\emph{Open problems.} First, Theorem~\ref{t:MC$+$sim_refined} improves the state bound; adapting faster majority~\cite{DBLP:conf/focs/DotyEGSUS21}, leader election, clocks, or conversions to MC$+$ could sharpen the bounds further. A structural goal is a state bound depending on memory and input/output requirements without a multiplicative dependence on running time, while preserving failure detection. Second, under the assumptions of Section~\ref{s:related}, a gap remains between $\poly(w)$-time source computations and reverse simulation in $\poly(w)$ space and expected $2^{O(w)}$ time. For predicates, the cited uniformity and constructibility assumptions also yield an $\mathsf{NSPACE}(\operatorname{polylog}n)$ upper bound~\cite{CzernerFischerGuttenberg25}. Narrowing this gap, characterising the power exactly, and determining whether dropping silence increases it remain open. Another goal is to remove advance knowledge of~$w$ while preserving arithmetic semantics and resource guarantees; stable size estimation~\cite{DBLP:conf/podc/BerenbrinkKR19} may help. Interaction graphs beyond the complete graph provide another direction.

\emph{External, distributed, and spatial data.} Relevant precedents include selective protocols for median computation using comparisons of arbitrarily large hidden keys~\cite{GanczorzGJKS24}, and spatial population protocols for localisation using distance or vector queries~\cite{GKL+24}. A useful goal is to integrate such interfaces with our simulation through operations for reading records, comparing keys, and acquiring geometric information that preserve anonymity and support efficient stable computation. Their costs must be analysed alongside internal arithmetic. Selection on distributed records and computation from statistical or geometric summaries provide concrete benchmarks. These extensions should account for precision, output representation, and data consistency, and support recovery followed by silence once the data stop changing.

\paragraph{AI methodology.}
OpenAI's ChatGPT and Anthropic's Claude assisted with language polishing, LaTeX editing, bibliographic reconciliation, and consistency checking. ChatGPT also helped refine the presentation of the simulation and recovery arguments in Sections~\ref{s:model}--\ref{s:stable_sim}, and Claude helped revise the introduction and related work, check cited claims against their sources, and review the manuscript. The authors verified the correctness and originality of all content, including references.

\clearpage

\bibliographystyle{wordram-alpha}
\bibliography{references}

\end{document}